\documentclass[aps,prd,showpacs,onecolumn,twoside]{revtex4-2}
\usepackage{amssymb}
\usepackage{mathrsfs}
\usepackage{amsfonts}
\usepackage{epstopdf}
\usepackage{bm,color,bbm}
\usepackage{amsmath}
\usepackage{graphicx}
\usepackage{natbib}
\usepackage[hidelinks]{hyperref}
\usepackage{caption}
\usepackage{subcaption}

\usepackage{fancyhdr}
\providecommand{\U}[1]{\protect\rule{.1in}{.1in}}

\begin{document}
\title{\textbf{Symmetry-Based Perspectives on Hamiltonian Quantum Search Algorithms
and Schr\"{o}dinger's Dynamics between Orthogonal States}}
\author{\textbf{Carlo Cafaro}$^{1}$ and \textbf{James Schneeloch}$^{2}$}
\affiliation{$^{1}$University at Albany-SUNY, Albany, NY 12222, USA}
\affiliation{$^{2}$Air Force Research Laboratory, Information Directorate, Rome, New York,
13441, USA}

\begin{abstract}
It is known that the continuous-time variant of Grover's search algorithm is
characterized by quantum search frameworks that are governed by stationary
Hamiltonians, which result in search trajectories confined to the
two-dimensional subspace of the complete Hilbert space formed by the source
and target states. Specifically, the search approach is ineffective when the
source and target states are orthogonal.

In this paper, we employ normalization, orthogonality, and energy limitations
to demonstrate that it is unfeasible to breach time-optimality between
orthogonal states with constant Hamiltonians when the evolution is limited to
the two-dimensional space spanned by the initial and final states. Deviations
from time-optimality for unitary evolutions between orthogonal states can only
occur with time-dependent Hamiltonian evolutions or, alternatively, with
constant Hamiltonian evolutions in higher-dimensional subspaces of the entire
Hilbert space. Ultimately, we employ our quantitative analysis to provide
meaningful insights regarding the relationship between time-optimal evolutions
and analog quantum search methods. We determine that the challenge of
transitioning between orthogonal states with a constant Hamiltonian in a
sub-optimal time is closely linked to the shortcomings of analog quantum
search when the source and target states are orthogonal and not interconnected
by the search Hamiltonian. In both scenarios, the fundamental cause of the
failure lies in the existence of an inherent symmetry within the system.

\end{abstract}
\maketitle

\affiliation{$^{1}$University at Albany-SUNY, Albany, NY 12222, USA}
\affiliation{$^{2}$Air Force Research Laboratory, Information Directorate, Rome, New York,
13441, USA}

\thispagestyle{fancy}

\section{Introduction}

Searching a marked item within an unstructured database by means of a
continuous-time evolution, characterized by a running time proportional to
$\sqrt{N}$, where $N$ represents the dimension of the search space, is a
significant endeavor in the realm of analog quantum computing
\cite{farhi98,carloijqi,steven20}. Additionally, a highly beneficial objective
in continuous-time quantum computation involves the development of Hamiltonian
evolutions that can transition the system from specified initial states to
final states in the shortest time possible \cite{ali09,bender07,cafarocqg}. A
key distinction between analog quantum search and optimal time evolutions lies
in the fact that, in the former scenario, the time evolution transpires
between a known source state and an unknown target state. Conversely, in the
latter scenario, the time evolution occurs between known initial and final
states. Nevertheless, a crucial element common to both analog quantum search
and optimal-time evolutions is the ability to restrict focus to the
two-dimensional subspace of the complete $N$-dimensional Hilbert space
\cite{farhi98,brody03}, which is defined by the source (initial) and target
(final) states, as the objective is to achieve optimal Hamiltonian evolutions.

\medskip

The initial continuous-time adaptation of Grover's original discrete quantum
search algorithm \cite{grover,grover05,zalka,wadati,cafaro17} was introduced
by Farhi and Gutmann in Ref. \cite{farhi98}. Afterwards, Bae and Kwon examined
a generalization of the Farhi-Gutmann analog quantum search algorithm in Ref.
\cite{bae02}. Both quantum search Hamiltonians discussed in Refs.
\cite{farhi98,bae02} are independent of time. The shift from time-independence
to time-dependence within the context of analog quantum search algorithms was
initially proposed under the assumption of global adiabatic evolution in Ref.
\cite{farhi00}, and subsequently, in a more favorable context of local
adiabaticity in Ref. \cite{roland02}. In the local adiabatic evolution
framework, the time-dependent search Hamiltonian is defined as a linear
interpolation between two time-independent Hamiltonians. In this scenario, the
system is initialized in the ground state of one of the two Hamiltonians. This
Hamiltonian is then adiabatically transformed into the other Hamiltonian,
whose ground state is presumed to represent the unknown solution to the search
problem. The time-dependence of the Hamiltonian is captured by the
time-dependent interpolation function. Notably, the specific form of this
function can be established by enforcing the local adiabaticity condition at
every moment during the quantum mechanical evolution of the system. In general
terms, the adiabatic approximation posits that a system prepared in an
instantaneous eigenstate of the Hamiltonian will remain close to this prepared
state if the Hamiltonian undergoes changes at a sufficiently slow rate
\cite{sanders04,oh05}. Consequently, any functional form of the interpolation
schedule that meets the requisite degree of slowness for the adiabatic
condition to be satisfied can be formally regarded. Interestingly, there is no
inherent reason to preclude a nonlinear interpolation of the two static
Hamiltonians \cite{ali04}. Indeed, Perez and Romanelli examined nonadiabatic
quantum search Hamiltonians that are defined through a nonlinear interpolation
of the two time-independent Hamiltonians referenced in Ref. \cite{romanelli07}%
. They introduced two search algorithms that display distinct characteristics,
depending on the parametric and temporal forms of the two interpolation
functions utilized to establish the time-dependent Hamiltonian. Notably, both
algorithms demonstrated the characteristic $\mathcal{O(}\sqrt{N})$ Grover-like
scaling behavior when locating the target state within an $N$-dimensional
search space. However, one of the algorithms necessitated a specific time for
measurement to identify the target state, exhibiting periodic oscillatory
behavior akin to the original Grover search algorithm
\cite{grover,cafaro12a,cafaro12b}. Conversely, the other algorithm led the
source state to evolve asymptotically into a quantum state that significantly
overlapped with the target state, showcasing monotonic behavior and resembling
the fixed-point Grover search algorithm \cite{grover05,cafaro17}.

\medskip

When examining the efficiency of quantum-mechanical unitary evolutions defined
by Hermitian Hamiltonians, under which an initial unit state vector
transitions to a final unit state vector \cite{anandan90,hamilton23}, two
primary alternative methodologies arise in the literature
\cite{ali09,bender07}. The first methodology, proposed by Mostafazadeh in Ref.
\cite{ali09}, involves seeking an expression for the Hamiltonian by maximizing
the energy uncertainty of the quantum system. This initial approach is
supported by the relationship between the angular velocity of the minimal-time
evolution of the quantum system and the energy uncertainty. Conversely, the
second methodology, introduced by Bender and colleagues in Ref.
\cite{bender07}, aims to derive an expression for the Hamiltonian by
minimizing the time required for the evolution from the initial to the final
state, while ensuring that the difference between the largest and smallest
eigenvalues of the Hamiltonian remains constant. Considering that the maximum
energy uncertainty correlates with the difference between the largest and
smallest energy eigenvalues, constraining the difference between the largest
and smallest eigenvalues of the Hamiltonian effectively constrains the energy
uncertainty. Consequently, it is reasonable to anticipate that these two
quantum descriptions of geodesic Hamiltonian motion are fundamentally
equivalent. Upon closer examination, it becomes evident that these two
methodologies emphasize slightly different characteristics. Notably, the
unique aspects of these two distinct methods for describing geodesic quantum
evolutions were skillfully utilized in Ref. \cite{cafaro22} to aid in
delineating the formal analogies between the geometry of quantum evolutions
exhibiting unit quantum geometric efficiency and the geometry of classical
polarization optics for light waves, where the degree of polarization matches
the degree of coherence between the electric vibrations in any two mutually
orthogonal directions of wave propagation \cite{wolf59,wolf07}. While these
distinguishing characteristics of Mostafazadeh's and Bender's methodologies
were briefly mentioned in Ref. \cite{cafaro22}, a comprehensive comparative
analysis of the two approaches found in Ref. \cite{ali09} and Ref.
\cite{bender07} is presented in Ref. \cite{rossetti24}.

\medskip

Symmetry arguments hold significant strength in various research domain
\cite{haywood}, including quantum walks \cite{lu16,chen17}, quantum transport
\cite{manzano18}, and quantum algorithms \cite{bottarelli25}. In Ref.
\cite{lu16}, for example, the authors present the concept of a chiral quantum
walk. This refers to a quantum walk on a graph or lattice where complex-valued
couplings are incorporated into the Hamiltonian to create directionality and
disrupt time-reversal symmetry. These complex phases afford enhanced control
over quantum transport, facilitating phenomena such as near-perfect state
transfer, directed excitation routing, and capabilities that exceed the
fundamental limits of traditional, non-chiral quantum walks. In Ref.
\cite{manzano18}, instead, the authors investigate the manner in which the
interaction between dissipative processes and an inherent symmetry propels the
quantum system of interest towards a degenerate steady state. This state
retains a portion of the information from the initial state, attributed to the
absence of mixing among the various symmetry sectors. Consequently, this paves
the way for comprehensive control over transport properties, including both
the average current and its statistical characteristics, by manipulating this
information through initial-state preparation methods. In Ref.
\cite{bottarelli25}, lastly, the authors introduce a symmetry-enhanced CD
(counterdiabatic)-inspired variational ansatz tailored for qudit systems. They
leverage the principle that the symmetries inherent in graph-defined problems
can enhance the efficacy of VQAs (variational quantum algorithms) while
simultaneously alleviating the demands on both classical and quantum hardware.
By employing these symmetries, the authors effectively minimize the
algorithm's resource requirements, particularly concerning the number of
parameters, which currently poses a significant limitation for VQAs. Their
ansatz strategically utilizes the system's symmetries to markedly decrease the
quantity of free parameters available for optimization.

\medskip

In the realm of analog quantum searching utilizing time-independent
Hamiltonians, the search strategy encounters failure when the source and
target states are orthogonal \cite{farhi98}. Similarly, in the context of
optimal-time evolutions, it is impossible to evolve in a two-dimensional space
in a time suboptimal manner between two orthogonal states using a stationary
Hamiltonian \cite{brody03}. Driven by these two shortcomings in the presence
of orthogonal states, this paper aims to enhance our comprehension of the
unique characteristics of continuous-time evolutions between orthogonal
quantum states. More specifically, a partial list of questions being explored includes:

\begin{enumerate}
\item[{[i]}] Why do stationary continuous-time quantum search Hamiltonians
fail to be effective when there is no quantum overlap between the source and
target states?

\item[{[ii]}] Do time-dependent search Hamiltonians mitigate work failures
such as, for instance, failure of the algorithm to reach the target state?
Alternatively, could the orthogonality between source and target states
continue to present a challenge that leads to work failure during the search?

\item[{[iii]}] What distinguishes quantum evolutions with stationary
Hamiltonians between orthogonal initial and final states in the
two-dimensional subspace defined by these states as particularly significant?

\item[{[iv]}] Do deviations from stationarity facilitate the development of
suboptimal Hamiltonian evolutions between orthogonal states?

\item[{[v]}] Is the failure to transition between orthogonal states with a
constant Hamiltonian in a sub-optimal time related to the ineffectiveness of
analog quantum search when the source and target states are orthogonal?
\end{enumerate}

\medskip

Properly addressing these questions is important because it can deepen our
theoretical understanding of the common foundations underlying both the
successes and failures of analog quantum searches and time-optimal quantum
evolutions, grounded in first physical principles and symmetry arguments.

\medskip

This rest of the paper is structured as follows. In Section II, we begin with
a discussion of the essential features of relevant continuous-time quantum
search Hamiltonians, followed by an outline of the main considerations from
the perspective of energy constraints when formulating optimal-time
Hamiltonians. In Section III, we employ normalization, orthogonality, and
energy constraints to demonstrate that it is unfeasible to breach
time-optimality between orthogonal states with constant Hamiltonians when the
evolution is restricted to the two-dimensional space defined by the initial
and final states. In Section IV, we present an analytical example that
illustrates the potential for deviating from time-optimal evolution between
orthogonal states within the two-dimensional subspace (of a larger Hilbert
space) that they generate, taking into account that the Hamiltonian may change
over time. In Section V, we offer physical insights into how one can avoid
analog quantum search failures by drawing parallels with occurrences in
standard quantum evolutions between specified initial and final states. Our
concluding observations are found in Section VI. Lastly, technical details
appear in Appendices A-C.

\section{Preliminaries}

In this section, following the discussion of the key characteristics of
pertinent continuous-time quantum search Hamiltonians, we outline the primary
considerations from an energy-constraints viewpoint when developing
optimal-time Hamiltonians.

\subsection{Analog quantum search}

Quantum search algorithms, such as Grover's quantum search scheme
\cite{grover,cafaro17}, were first introduced within a digital quantum
computation framework characterized by a discrete series of unitary logic
gates. In contrast, Farhi and Gutmann employed an analog quantum computation
approach to introduce an analog variant of Grover's original quantum search
algorithm, wherein the quantum register's state experiences a continuous time
evolution influenced by a suitably chosen driving Hamiltonian \cite{farhi98}.
The fundamental concept of the continuous time search algorithm introduced by
Farhi and Gutmann can be encapsulated as follows. Given a Hamiltonian that
operates on an $N$-dimensional (where $N=2^{n}$) complex vector space
$\mathcal{H}_{2}^{n}$, which possesses a single non-zero eigenvalue $E\neq0$
while all other eigenvalues are zero, the objective is to identify the
eigenvector $\left\vert w\right\rangle $ corresponding to the eigenvalue $E$.
The search concludes when the quantum system is confirmed to be in the state
$\left\vert w\right\rangle $. By utilizing time-independent Hamiltonian
evolutions, Farhi and Gutmann demonstrated that their algorithm necessitated a
minimum search duration on the order of $\sqrt{N}$, thereby exhibiting the
same complexity as Grover's original quantum search algorithm. The complete
original Farhi-Gutmann quantum search Hamiltonian as detailed in
\cite{farhi98} is given by
\begin{equation}
\mathrm{H}_{\text{\textrm{FG}}}\overset{\text{def}}{=}\mathrm{H}%
_{w}+\mathrm{H}_{d}=E\left\vert w\right\rangle \left\langle w\right\vert
+E\left\vert s\right\rangle \left\langle s\right\vert \text{,} \label{FG}%
\end{equation}
where $\mathrm{H}_{w}\overset{\text{def}}{=}E\left\vert w\right\rangle
\left\langle w\right\vert $ and $\mathrm{H}_{d}\overset{\text{def}}%
{=}E\left\vert s\right\rangle \left\langle s\right\vert $ denote the oracle
and driving Hamiltonians, respectively. The normalized states $\left\vert
s\right\rangle $ and $\left\vert w\right\rangle $ represent the initial
(source) and final (target) states, respectively. It is important to note
that, in order to comprehend the analog search process, it is adequate to
concentrate on the two-dimensional subspace formed by $\left\vert
s\right\rangle $ and $\left\vert w\right\rangle $. The target state
$\left\vert w\right\rangle $ is an arbitrarily selected (unknown) state from
the unit sphere in $\mathcal{H}_{2}^{n}$, whereas the source state $\left\vert
s\right\rangle $ is a suitably chosen normalized vector that is independent of
$\left\vert w\right\rangle $. The source state $\left\vert s\right\rangle $
evolves in accordance with Schr\"{o}dinger's evolution law, $\left\vert
s\right\rangle \mapsto\left\vert \psi\left(  t\right)  \right\rangle
\overset{\text{def}}{=}e^{-\frac{i}{\hslash}\mathrm{H}_{\text{FG}}t}\left\vert
s\right\rangle $. Furthermore, with no loss of generality, the quantum overlap
$x\overset{\text{def}}{=}\left\langle w|s\right\rangle \neq0$ can be assumed
to be real and positive since any phase factor in the inner product between
these two states can be ultimately inserted into the state $\left\vert
s\right\rangle $. In addition, since it suffices to limit our attention to the
two-dimensional subspace of $\mathcal{H}_{2}^{n}$ spanned by $\left\vert
s\right\rangle $ and $\left\vert w\right\rangle $, it is useful to introduce
the orthonormal basis\textbf{\ }$\left\{  \left\vert w\right\rangle \text{,
}\left\vert r\right\rangle \right\}  $ with $\left\vert r\right\rangle
\overset{\text{def}}{=}\left(  1-x^{2}\right)  ^{-1/2}\left(  \left\vert
s\right\rangle -x\left\vert w\right\rangle \right)  $ and\textbf{ }$\left\vert
s\right\rangle =x\left\vert w\right\rangle +\sqrt{1-x^{2}}\left\vert
r\right\rangle $, respectively. Making use of the basis $\left\{  \left\vert
w\right\rangle \text{, }\left\vert r\right\rangle \right\}  $, it happens that
the probability $\mathcal{P}_{\text{\textrm{FG}}}\left(  t\right)  $ of
finding the state $\left\vert w\right\rangle $ at time time $t$ reduces to
\cite{farhi98},%
\begin{equation}
\mathcal{P}_{\text{\textrm{FG}}}\left(  t\right)  \overset{\text{def}}%
{=}\left\vert \left\langle w|e^{-\frac{i}{\hslash}\mathrm{H}%
_{\text{\textrm{FG}}}t}|s\right\rangle \right\vert ^{2}=\sin^{2}\left(
\frac{Ex}{\hslash}t\right)  +x^{2}\cos^{2}\left(  \frac{Ex}{\hslash}t\right)
\text{.} \label{PFG}%
\end{equation}
\textbf{\ }Specifically, the earliest moment $t_{\text{\textrm{FG}}}$ when the
transition probability $\mathcal{P}_{\text{\textrm{FG}}}\left(  t\right)  $ in
Eq. (\ref{PFG}) assumes its maximum value $\mathcal{P}_{\text{\textrm{FG}}%
}^{\max}=1$ is given by,%
\begin{equation}
t_{\text{\textrm{FG}}}\overset{\text{def}}{=}\frac{\hslash}{E}\frac{\pi}%
{2x}\text{.} \label{tgf}%
\end{equation}
For later use, we emphasize that the energy dispersion of $\mathrm{H}%
_{\text{\textrm{FG}}}$ evaluated with respect to to source state $\left\vert
s\right\rangle $ is given by $\Delta E_{\text{\textrm{FG}}}\overset
{\text{def}}{=}Ex\sqrt{1-x^{2}}$. When the target state $\left\vert
w\right\rangle $ is considered to be a member of a collection of mutually
orthonormal quantum states $\left\{  \left\vert a\right\rangle \right\}  $
$1\leq a\leq N$ in $\mathcal{H}_{2}^{n}$, the source state $\left\vert
s\right\rangle $ can be effectively selected as an equal superposition of the
$N$ quantum states $\left\{  \left\vert a\right\rangle \right\}  $.
Consequently, $x=1/\sqrt{N}$ and from Eq. (\ref{tgf}) we observe that
$t_{\text{\textrm{FG}}}\propto\sqrt{N}$. Therefore, similar to Grover's
search, the Farhi-Gutmann algorithm necessitates a minimum search duration on
the order of $\sqrt{N}$. Furthermore, by positing that the target state is an
unknown member of a specified orthonormal basis $\left\{  \left\vert
a\right\rangle \right\}  $ with $1\leq a\leq N$ in $\mathcal{H}_{2}^{n}$ that
is generated with complete certainty, Farhi and Gutmann demonstrated that
their algorithm is optimally short.

Despite the optimality of the Farhi-Gutmann quantum search Hamiltonian in Eq.
(\ref{FG}) (in the sense that $t_{\text{\textrm{FG}}}\propto\sqrt{N}\simeq
t_{\mathrm{Grover}}$), Fenner pointed out in Ref. \cite{fenner2000} that the
unitary evolution operator $U_{\mathrm{FG}}\left(  t\right)  =e^{-\frac
{i}{\hslash}\mathrm{H}_{\mathrm{FG}}t}$ deviates significantly from the
intermediate states attained in the original algorithm when Grover's iterate
is applied. Furthermore, he demonstrated that it is possible to consider a
different search Hamiltonian that can more accurately align with the actual
iterations defining Grover's algorithm. Fenner's search Hamiltonian is defined
as,%
\begin{equation}
\mathrm{H}_{\mathrm{Fenner}}\overset{\text{def}}{=}\frac{2i}{E}\left[
\mathrm{H}_{w}\text{, }\mathrm{H}_{d}\right]  =2iEx\left(  \left\vert
w\right\rangle \left\langle s\right\vert -\left\vert s\right\rangle
\left\langle w\right\vert \right)  \text{,} \label{Fenner}%
\end{equation}
where $\mathrm{H}_{w}$, $\mathrm{H}_{d}$, and $x$ are the same quantities
introduced for $\mathrm{H}_{\text{\textrm{FG}}}$ in Eq. (\ref{Fenner}). After
some algebra (for details, see Appendix A), it turns out that the analogue of
the transition probability in Eq. (\ref{PFG}) is given by
\begin{equation}
\mathcal{P}_{\text{\textrm{Fenner}}}\left(  t\right)  \overset{\text{def}}%
{=}\left\vert \left\langle w|e^{-\frac{i}{\hslash}\mathrm{H}%
_{\text{\textrm{Fenner}}}t}|s\right\rangle \right\vert ^{2}=\left\vert
x\cos\left(  2x\sqrt{1-x^{2}}\frac{E}{\hslash}t\right)  +\sqrt{1-x^{2}}%
\sin\left(  2x\sqrt{1-x^{2}}\frac{E}{\hslash}t\right)  \right\vert
^{2}\text{.} \label{PFenner}%
\end{equation}
\textbf{\ }In particular, the shortest time $t_{\text{\textrm{Fenner}}}$ when
the transition probability $\mathcal{P}_{\text{\textrm{Fenner}}}\left(
t\right)  $ in Eq. (\ref{PFenner}) assumes its maximum value $\mathcal{P}%
_{\text{\textrm{Fenner}}}^{\max}=1$ is given by,%
\begin{equation}
t_{\text{\textrm{Fenner}}}\overset{\text{def}}{=}\frac{\hslash}{E}\frac{\text{
}\arccos(x)}{2x\sqrt{1-x^{2}}}\text{.} \label{tgf2}%
\end{equation}
Interestingly, we point out that the energy dispersion of $\mathrm{H}%
_{\text{\textrm{Fenner}}}$ evaluated with respect to to source state
$\left\vert s\right\rangle $ is given by $\Delta E_{\text{\textrm{Fenner}}%
}\overset{\text{def}}{=}2Ex\sqrt{1-x^{2}}$. Therefore, keeping in mind that
$\Delta E_{\text{\textrm{Fenner}}}=2\Delta E_{\text{\textrm{FG}}}$, an
energetically fair comparison of Eqs. (\ref{tgf}) and (\ref{tgf2}) yields
$t_{\text{\textrm{FG}}}\simeq t_{\text{\textrm{Fenner}}}$ in the limiting case
in which $x\ll1$.

After covering the fundamentals of (time-independent) analog quantum search
Hamiltonians, we are now prepared to introduce the key aspects of optimal-time Hamiltonians.

\subsection{Optimal-time evolutions}

Following Refs. \cite{ali09,cafarocqg}, let us assume a traceless and
stationary Hamiltonian \textrm{H }defined by a spectral decomposition of the
form \textrm{H}$\overset{\text{def}}{=}E_{1}\left\vert E_{1}\right\rangle
\left\langle E_{1}\right\vert +E_{2}\left\vert E_{2}\right\rangle \left\langle
E_{2}\right\vert $, with $\left\langle E_{1}|E_{2}\right\rangle =\delta_{12}$
and $E_{2}\geq E_{1}$. For clarity, we note that $\left\{  \left\vert
E_{i}\right\rangle \right\}  _{i=1,2}$ and $\left\{  E_{i}\right\}  _{i=1,2}%
$\textbf{\ }specify the eigenvectors and the eigenvalues of the constant
Hamiltonian \textrm{H}. Furthermore, $\delta_{ij}$\textbf{\ }denotes the
Kronecker delta symbol, where\textbf{\ }$1\leq i$\textbf{, }$j\leq2$. In
time-optimal scenarios, one is interested in the evolution of a state
$\left\vert A\right\rangle $, not necessarily normalized, into a state
$\left\vert B\right\rangle $ in the shortest possible time by maximizing the
energy uncertainty $\Delta E$,
\begin{equation}
\Delta E\overset{\text{def}}{=}\left[  \frac{\left\langle A|\mathrm{H}%
^{2}\mathrm{|}A\right\rangle }{\left\langle A|A\right\rangle }-\left(
\frac{\left\langle A|\mathrm{H|}A\right\rangle }{\left\langle A|A\right\rangle
}\right)  ^{2}\right]  ^{1/2}\text{,} \label{deltaE}%
\end{equation}
so that $\Delta E$ in Eq. (\ref{deltaE}) equals $\Delta E_{\max}$. The reason
why one maximizes the energy uncertainty\textbf{\ }$\Delta E$\textbf{\ }is
justified by the fact that the speed of quantum evolution\textbf{\ }%
$ds/dt$\textbf{\ }along the curve that connects\textbf{\ }$\left\vert
A\right\rangle $ to $\left\vert B\right\rangle $\textbf{\ }is proportional to
the energy uncertainty\textbf{\ }$\Delta E$\textbf{, }$ds/dt\propto\Delta
E$\textbf{.} What is the value of $\Delta E_{\max}?$ To obtain this value, we
notice that any unnormalized initial state $\left\vert A\right\rangle $ can be
decomposed as $\left\vert A\right\rangle =\alpha_{1}\left\vert E_{1}%
\right\rangle +\alpha_{2}\left\vert E_{2}\right\rangle $ with generally
complex quantum amplitudes $\alpha_{1}\overset{\text{def}}{=}$ $\left\langle
E_{1}|A\right\rangle $ and $\alpha_{2}\overset{\text{def}}{=}\left\langle
E_{2}|A\right\rangle $. Using this decomposition of $\left\vert A\right\rangle
$ into Eq. (\ref{deltaE}), we arrive at%
\begin{equation}
\Delta E=\frac{E_{2}-E_{1}}{2}\left[  1-\left(  \frac{\left\vert \alpha
_{1}\right\vert ^{2}-\left\vert \alpha_{2}\right\vert ^{2}}{\left\vert
\alpha_{1}\right\vert ^{2}+\left\vert \alpha_{2}\right\vert ^{2}}\right)
^{2}\right]  ^{1/2}\text{.} \label{chi4}%
\end{equation}
Inspection of Eq. (\ref{chi4}) implies that the maximum value of $\Delta E$ in
Eq. (\ref{chi4}) is achieved when $\left\vert \alpha_{1}\right\vert
=\left\vert \alpha_{2}\right\vert $. In particular, it equals
\begin{equation}
\Delta E_{\max}\overset{\text{def}}{=}\left(  \frac{E_{2}-E_{1}}{2}\right)
\text{.}%
\end{equation}
A fundamental concept in Mostafazadeh's methodology as presented in Ref.
\cite{ali09} involves the expression of \textrm{H}$\overset{\text{def}}%
{=}E_{1}\left\vert E_{1}\right\rangle \left\langle E_{1}\right\vert
+E_{2}\left\vert E_{2}\right\rangle \left\langle E_{2}\right\vert $ in terms
of the initial and final states $\left\vert A\right\rangle $ and $\left\vert
B\right\rangle $, respectively, while ensuring that $\Delta E=\Delta E_{\max}%
$. In this context, it is important to recognize that $\left\vert
A\right\rangle $ and $\left\vert B\right\rangle $ can be expressed as
$\left\vert A\right\rangle =\alpha_{1}\left\vert E_{1}\right\rangle
+\alpha_{2}\left\vert E_{2}\right\rangle $ and $\left\vert B\right\rangle
=\beta_{1}\left\vert E_{1}\right\rangle +\beta_{2}\left\vert E_{2}%
\right\rangle $, respectively. Furthermore, it is necessary to impose
$\left\vert \alpha_{1}\right\vert =\left\vert \alpha_{2}\right\vert $ and
$\left\vert \beta_{1}\right\vert =\left\vert \beta_{2}\right\vert $ or,
alternatively,%
\begin{equation}
\left\vert \alpha_{2}\right\vert ^{2}-\left\vert \alpha_{1}\right\vert
^{2}=0=\left\vert \beta_{2}\right\vert ^{2}-\left\vert \beta_{1}\right\vert
^{2}\text{,} \label{bruno1}%
\end{equation}
in order to meet the condition $\Delta E=\Delta E_{\max}$ and thereby ensure
the minimum travel time $t_{AB}^{\min}\overset{\text{def}}{=}\hslash
\arccos\left(  \left\vert \left\langle A\left\vert B\right.  \right\rangle
\right\vert \right)  /\Delta E_{\max}$. After some straightforward but tedious
algebra, one can show that the expression of the time-optimal Hamiltonian
\textrm{H} that connects $\left\vert A\right\rangle $ to $\left\vert
B\right\rangle $ (assuming energy dispersion $\Delta E=E$, with $E_{2}%
=-E_{1}\overset{\text{def}}{=}E$) is given by%
\begin{equation}
\mathrm{H}_{\mathrm{opt}}\overset{\text{def}}{=}i\Delta E\frac{\left\vert
\left\langle A\left\vert B\right.  \right\rangle \right\vert }{\sqrt
{1-\left\vert \left\langle A\left\vert B\right.  \right\rangle \right\vert
^{2}}}\left[  \frac{\left\vert B\right\rangle \left\langle A\right\vert
}{\left\langle A|B\right\rangle }-\frac{\left\vert A\right\rangle \left\langle
B\right\vert }{\left\langle B|A\right\rangle }\right]  \text{.} \label{amy}%
\end{equation}
Recalling that quantum states that vary by solely a global phase are
physically equivalent, we stress for completeness the unitary time propagator
$U_{\mathrm{opt}}\left(  t\right)  \overset{\text{def}}{=}e^{-\frac{i}%
{\hslash}\mathrm{H}_{\mathrm{opt}}t}$ is such that $U_{\mathrm{opt}%
}(t_{\text{\textrm{opt}}})\left\vert A\right\rangle =\left\vert B\right\rangle
$ with $t_{\text{\textrm{opt}}}=t_{AB}^{\min}$ given by%
\begin{equation}
t_{\text{\textrm{opt}}}\overset{\text{def}}{=}\frac{\hslash\arccos\left(
\left\vert \left\langle A\left\vert B\right.  \right\rangle \right\vert
\right)  }{\Delta E}\text{.} \label{amy2}%
\end{equation}
Finally, we remark that for $\mathrm{H}_{\mathrm{opt}}$ in Eq. (\ref{amy}),
one correctly obtains $\left\langle A|\mathrm{H}_{\mathrm{opt}}\mathrm{|}%
A\right\rangle /\left\langle A|A\right\rangle =0$ and $\Delta E=\left[
\left\langle A|\mathrm{H}_{\mathrm{opt}}^{2}|A\right\rangle /\left\langle
A|A\right\rangle \right]  ^{1/2}=E=\Delta E_{\max}$.

From a time-evolution standpoint, we stress that the word \textquotedblleft%
\emph{optimality}\textquotedblright\ means that the duration of the evolution
from known initial and final states happens in an optimal time given by
$t_{\mathrm{opt}}^{\mathrm{evolution}}=\hslash\arccos\left[  \left\vert
\left\langle A\left\vert B\right.  \right\rangle \right\vert \right]  /\Delta
E$. From a quantum search viewpoint, instead, \textquotedblleft%
\emph{optimality}\textquotedblright\ signifies that producing an unknown
target state $\left\vert w\right\rangle $ starting from a known source state
$\left\vert s\right\rangle $ happens in a temporal duration that is within a
constant \textrm{c}$\in%
\mathbb{R}
_{+}\backslash\left\{  0\right\}  $ of the best possible. In other words,
$t_{\mathrm{opt}}^{\mathrm{search}}\propto\sqrt{N}$, with $N$ denoting the
dimensionality of the search space. Interestingly, the time evolution
sub-optimality of the Fahri-Gutmann search Hamiltonian can also be understood
in geometric terms by re-deriving the search Hamiltonian from the point of
view of quantum simulation (for details, see Appendix B). \ \textrm{H}%
$_{\mathrm{FG}}$ is time suboptimal and, therefore, cannot work for orthogonal
states. \textrm{H}$_{\mathrm{Fenner}}$, instead, is time optimal. However, it
does not work for orthogonal states by construction.

Having addressed the basics of constructing optimal-time Hamiltonians, we are
now prepared to explore deviations from time-optimality within two-dimensional subspaces.

\section{Our verification}

In this section, we utilize normalization, orthogonality, and energy
constraints to illustrate that it is impractical to violate time-optimality
between orthogonal states with constant Hamiltonians when the evolution is
confined to the two-dimensional space \ \ spanned by the initial and final states.

Prior to outlining our energy-based arguments, we will briefly examine the
geometric reasoning put forth by Brody in Ref. \cite{brody03}.

\subsection{Geometry-based reasoning}

What is the peculiarity of quantum evolutions between orthogonal quantum
states in two-dimensional subspaces of the full Hilbert space when the unitary
time propagator is specified by a time-independent Hamiltonian?

Following the analysis in Ref. \cite{brody03}, the peculiarity is that if one
assumes two-dimensional motion with a constant Hamiltonian, the only way to
evolve between orthogonal initial and final unit states $\left\vert
A\right\rangle $ and $\left\vert B\right\rangle $ is along a time-optimal
geodesic path. Why does this happen? First, since the states are orthogonal,
they are antipodal on the two-sphere.\ Second, if $\left\vert A\right\rangle $
is a point on the projective line specified by a pair of eigenstates $\left(
\left\vert E_{+}\right\rangle \text{, }\left\vert E_{-}\right\rangle \right)
$ of the Hamiltonian \textrm{H}, then the trajectory $\gamma\left(  t\right)
$ given by
\begin{equation}
\gamma\left(  t\right)  :t\mapsto\left\vert \psi\left(  t\right)
\right\rangle \overset{\text{def}}{=}U(t)\left\vert A\right\rangle
=e^{-\frac{i}{\hslash}\mathrm{H}t}\left\vert A\right\rangle \label{U2}%
\end{equation}
never leaves the above mentioned projective line. Therefore, the length of the
path between the antipodal states $\left\vert A\right\rangle $ and $\left\vert
B\right\rangle \overset{\text{def}}{=}\left\vert \psi\left(  t_{\mathrm{final}%
}\right)  \right\rangle =U(t_{\mathrm{final}})\left\vert A\right\rangle $ is
$\pi$. Indeed, the (geodesic) length $\mathcal{L}\left[  \gamma\left(
t\right)  \right]  _{0\leq t\leq t_{\mathrm{final}}}$ of the path
$\gamma\left(  t\right)  $ in Eq. (\ref{U2}) with $0\leq t\leq
t_{\mathrm{final}}$ as measured by the Fubini-Study metric is given by (for
details, see Appendix C)
\begin{equation}
\mathcal{L}\left[  \gamma\left(  t\right)  \right]  _{0\leq t\leq
t_{\mathrm{final}}}=2\arccos\left[  \left\vert \left\langle A\left\vert
B\right.  \right\rangle \right\vert \right]  \text{,} \label{lunghezza}%
\end{equation}
where $\left\vert A\right\rangle \overset{\text{def}}{=}\left\vert \psi\left(
0\right)  \right\rangle $ and $\left\vert B\right\rangle \overset{\text{def}%
}{=}\left\vert \psi\left(  t_{\mathrm{final}}\right)  \right\rangle $
\cite{cafarocqg}. Then, since the path length is $\pi$ for any pair of energy
eigenstates $\left(  \left\vert E_{+}\right\rangle \text{, }\left\vert
E_{-}\right\rangle \right)  $ that may specify other constant Hamiltonians,
there exists no alternative path with a length longer than $\pi$. As a side
note, we wish to clarify that the projective line connecting two points
$\left\vert P_{1}\right\rangle $ and $\left\vert P_{2}\right\rangle $ is
defined as the line whose points signify the entirety of normalized
superpositions of the states $\left\vert P_{1}\right\rangle $ and $\left\vert
P_{2}\right\rangle $. With a slight deviation from strict terminology, we
interchangeably use the terms \textquotedblleft points\textquotedblright\ and
\textquotedblleft states\textquotedblright. Nevertheless, it is crucial to
remember that these \textquotedblleft points\textquotedblright\ are
constituents of the projective Hilbert space $\mathcal{P(H)}$ (i.e., the space
of rays) equipped with the Fubini-Study metric. The concept of a
\textquotedblleft ray\textquotedblright\ as an equivalence class of state
vectors in the Hilbert space $\mathcal{H}$ addresses the issue that global
phase factors are physically insignificant, resulting in $\mathcal{H}$
possessing an inherently redundant complex degree of freedom. For further
information, we direct the reader to \cite{brody03}. Therefore, when regarded
as points on the projective line specified by $\left(  \left\vert
E_{+}\right\rangle \text{, }\left\vert E_{-}\right\rangle \right)  $, the unit
state vectors $\left\vert A\right\rangle $ and $\left\vert B\right\rangle $
can be decomposed as
\begin{equation}
\left\vert A\right\rangle =\left\langle E_{+}\left\vert A\right.
\right\rangle \left\vert E_{+}\right\rangle +\left\langle E_{-}\left\vert
A\right.  \right\rangle \left\vert E_{-}\right\rangle \text{, and }\left\vert
B\right\rangle =\left\langle E_{+}\left\vert B\right.  \right\rangle
\left\vert E_{+}\right\rangle +\left\langle E_{-}\left\vert B\right.
\right\rangle \left\vert E_{-}\right\rangle \text{,}%
\end{equation}
respectively. In particular, normalization of $\left\vert A\right\rangle $ and
$\left\vert B\right\rangle $ implies $\left\vert \left\langle E_{+}\left\vert
A\right.  \right\rangle \right\vert ^{2}+\left\vert \left\langle
E_{-}\left\vert A\right.  \right\rangle \right\vert ^{2}=1$ and $\left\vert
\left\langle E_{+}\left\vert B\right.  \right\rangle \right\vert
^{2}+\left\vert \left\langle E_{-}\left\vert B\right.  \right\rangle
\right\vert ^{2}=1$, respectively. In particular, when $\left\vert
A\right\rangle $ and $\left\vert B\right\rangle $ are orthogonal and, in
addition, \textrm{H}$=E_{+}\left\vert E_{+}\right\rangle \left\langle
E_{+}\right\vert +E_{-}\left\vert E_{-}\right\rangle \left\langle
E_{-}\right\vert $ is stationary, the only possible scenario is $\left\vert
\left\langle E_{+}\left\vert A\right.  \right\rangle \right\vert
^{2}=\left\vert \left\langle E_{-}\left\vert A\right.  \right\rangle
\right\vert ^{2}=1/2$ and $\left\vert \left\langle E_{+}\left\vert B\right.
\right\rangle \right\vert ^{2}=\left\vert \left\langle E_{-}\left\vert
B\right.  \right\rangle \right\vert ^{2}=1/2$.

In conclusion, the distinctive feature of orthogonal states $\left\vert
A\right\rangle $ and $\left\vert B\right\rangle $ is that their relative
distance, defined by the Fubini-Study metric, consistently measures $\pi$
(irrespective of the pair of energy eigenstates one may choose to examine).
Conversely, for nonorthogonal states $\left\vert A\right\rangle $ and
$\left\vert B\right\rangle $, it is generally observed that the relative
distance between them on different projective lines (determined by appropriate
pairs of energy eigenstates linked to specific stationary Hamiltonians) tends
to vary. For straightforward illustrative examples displaying optimal and
suboptimal Hamiltonian evolutions between nonorthogonal quantum states on the
Bloch sphere, we recommend consulting Ref. \cite{carloepjplus}.

Following a concise examination of the geometric rationale put forth by Brody
to elucidate the absence of paths exceeding $\pi$ in length between orthogonal
states, when the evolution is limited to the two-dimensional subspace they
span, we are now ready to present our alternative perspective based on energy considerations.

\subsection{Energy-based reasoning}

In the following, we present our alternative approach to clarify the lack of
paths longer than $\pi$ in length that connect orthogonal states, particularly
when the evolution is confined to the two-dimensional subspace they occupy.

Let us assume an arbitrary parametrization of the unit state vectors
$\left\vert A\right\rangle $ and $\left\vert B\right\rangle $ given by,%
\begin{equation}
\left\vert A\right\rangle =\alpha_{1}\left\vert E_{1}\right\rangle +\alpha
_{2}\left\vert E_{2}\right\rangle \text{, and }\left\vert B\right\rangle
=\beta_{1}\left\vert E_{1}\right\rangle +\beta_{2}\left\vert E_{2}%
\right\rangle \text{,} \label{c0}%
\end{equation}
respectively, where $\alpha_{1}$, $\alpha_{2}$, $\beta_{1}$, and $\beta_{2}$
are complex quantum amplitudes. In particular, we assume in Eq. (\ref{c0})
that the spectral decomposition of the Hamiltonian restricted to the
two-dimensional subspace spanned by $\left\vert A\right\rangle $ and
$\left\vert B\right\rangle $ is given by \textrm{H}$=E_{1}\left\vert
E_{1}\right\rangle \left\langle E_{1}\right\vert +E_{2}\left\vert
E_{2}\right\rangle \left\langle E_{2}\right\vert $. The states $\left\vert
A\right\rangle $ and $\left\vert B\right\rangle $ have to satisfy both
normalization (i.e., $\left\langle A\left\vert A\right.  \right\rangle
=1=\left\langle B\left\vert B\right.  \right\rangle $) and orthogonality
conditions (i.e., $\left\langle A\left\vert B\right.  \right\rangle
=0=\left\langle B\left\vert A\right.  \right\rangle $). Therefore, it must be%
\begin{equation}
\left\vert \alpha_{1}\right\vert ^{2}+\left\vert \alpha_{2}\right\vert
^{2}=1=\left\vert \beta_{1}\right\vert ^{2}+\left\vert \beta_{2}\right\vert
^{2}\text{,} \label{c1}%
\end{equation}
and%
\begin{equation}
\alpha_{1}^{\ast}\beta_{1}+\alpha_{2}^{\ast}\beta_{2}=0=\alpha_{1}\beta
_{1}^{\ast}+\alpha_{2}\beta_{2}^{\ast}\text{,} \label{c2}%
\end{equation}
respectively. Most importantly, since we want to connect $\left\vert
A\right\rangle $ to $\left\vert B\right\rangle $ via a unitary time-propagator
$U(t)=e^{-\frac{i}{\hslash}\mathrm{H}t}$, there must exist an instant
$t_{\mathrm{final}}$ such that $\left\vert B\right\rangle =e^{-\frac
{i}{\hslash}\mathrm{H}t_{\mathrm{final}}}\left\vert A\right\rangle $.
Therefore, given that $\left\vert B\right\rangle =e^{-\frac{i}{\hslash
}\mathrm{H}t_{\mathrm{final}}}\left\vert A\right\rangle $ with \textrm{H}
time-independent, it must be%
\begin{equation}
\left\langle A\left\vert \mathrm{H}\right\vert A\right\rangle =\left\langle
B\left\vert \mathrm{H}\right\vert B\right\rangle \text{,} \label{c3}%
\end{equation}
and, in particular,%
\begin{equation}
\Delta E_{A}^{2}\overset{\text{def}}{=}\left\langle A\left\vert \mathrm{H}%
^{2}\right\vert A\right\rangle -\left\langle A\left\vert \mathrm{H}\right\vert
A\right\rangle ^{2}=\left\langle B\left\vert \mathrm{H}^{2}\right\vert
B\right\rangle -\left\langle B\left\vert \mathrm{H}\right\vert B\right\rangle
^{2}\overset{\text{def}}{=}\Delta E_{B}^{2}\text{.} \label{c4}%
\end{equation}
The condition on the mean energy in Eq. (\ref{c3}) implies%
\begin{equation}
E\left(  \left\vert \alpha_{2}\right\vert ^{2}-\left\vert \alpha
_{1}\right\vert ^{2}\right)  =E\left(  \left\vert \beta_{2}\right\vert
^{2}-\left\vert \beta_{1}\right\vert ^{2}\right)  \text{,}%
\end{equation}
that is,%
\begin{equation}
\left\vert \alpha_{2}\right\vert ^{2}-\left\vert \alpha_{1}\right\vert
^{2}=\left\vert \beta_{2}\right\vert ^{2}-\left\vert \beta_{1}\right\vert
^{2}\text{.} \label{c5}%
\end{equation}
Moreover, the condition on the energy variance in Eq. (\ref{c4}) leads to%
\begin{equation}
E^{2}\left[  1-\left(  \left\vert \alpha_{2}\right\vert ^{2}-\left\vert
\alpha_{1}\right\vert ^{2}\right)  ^{2}\right]  =E^{2}\left[  1-\left(
\left\vert \beta_{2}\right\vert ^{2}-\left\vert \beta_{1}\right\vert
^{2}\right)  ^{2}\right]  \text{,}%
\end{equation}
that is,%
\begin{equation}
\left(  \left\vert \alpha_{2}\right\vert ^{2}-\left\vert \alpha_{1}\right\vert
^{2}\right)  ^{2}=\left(  \left\vert \beta_{2}\right\vert ^{2}-\left\vert
\beta_{1}\right\vert ^{2}\right)  ^{2}\text{.} \label{c6}%
\end{equation}
We observe that the condition in Eq. (\ref{c5}) implies the condition in Eq.
(\ref{c6}) (but not vice versa). Therefore, in summary, we have that the
complex amplitudes $\alpha_{1}$, $\alpha_{2}$, $\beta_{1}$, and $\beta_{2}$
must simultaneously satisfy the following set of constraints%
\begin{equation}
\left\{
\begin{array}
[c]{c}%
\left\vert \alpha_{1}\right\vert ^{2}+\left\vert \alpha_{2}\right\vert
^{2}=1=\left\vert \beta_{1}\right\vert ^{2}+\left\vert \beta_{2}\right\vert
^{2}\\
\left\vert \alpha_{2}\right\vert ^{2}-\left\vert \alpha_{1}\right\vert
^{2}=\left\vert \beta_{2}\right\vert ^{2}-\left\vert \beta_{1}\right\vert
^{2}\\
\alpha_{1}^{\ast}\beta_{1}+\alpha_{2}^{\ast}\beta_{2}=0=\alpha_{1}\beta
_{1}^{\ast}+\alpha_{2}\beta_{2}^{\ast}%
\end{array}
\right.  \text{.} \label{c7}%
\end{equation}
Without loss of generality, we can set
\begin{equation}
\left\vert \alpha_{2}\right\vert ^{2}-\left\vert \alpha_{1}\right\vert
^{2}=\varepsilon=\left\vert \beta_{2}\right\vert ^{2}-\left\vert \beta
_{1}\right\vert ^{2}\text{,} \label{amy3}%
\end{equation}
with $\varepsilon\in%
\mathbb{R}
$ at this stage. We stress that the conditions in Eq. (\ref{amy3}) represent
the key departure from what is supposed to happen in a time-optimal setting as
summarized in Eq. (\ref{bruno1}). Then, using the fact that $\left\vert
\alpha_{1}\right\vert ^{2}+\left\vert \alpha_{2}\right\vert ^{2}=1=\left\vert
\beta_{1}\right\vert ^{2}+\left\vert \beta_{2}\right\vert ^{2}$, we get from
the first two conditions in Eq. (\ref{c7}) that%
\begin{equation}
\left\vert \alpha_{1}\right\vert ^{2}=\frac{1-\varepsilon}{2}\text{,
}\left\vert \alpha_{2}\right\vert ^{2}=\frac{1+\varepsilon}{2}\text{,
}\left\vert \beta_{1}\right\vert ^{2}=\frac{1-\varepsilon}{2}\text{, and
}\left\vert \beta_{2}\right\vert ^{2}=\frac{1+\varepsilon}{2}\text{,}
\label{c8}%
\end{equation}
with $\varepsilon\in\left(  0\text{, }1\right)  $ (given the positivity of
both $\left\vert \alpha_{1}\right\vert ^{2}$ and $\left\vert \beta
_{1}\right\vert ^{2}$). As a side remark, we observe that when $\varepsilon
=0$, we recover the time-optimal scenario parametrization of the states
$\left\vert A\right\rangle $ and $\left\vert B\right\rangle $ in Eq.
(\ref{c0}). Returning to our discussion, making use of Eqs. (\ref{c0}) and
(\ref{c8}), we can recast $\left\vert A\right\rangle $ and $\left\vert
B\right\rangle $ in Eq. (\ref{c0}) as
\begin{equation}
\left\vert A\right\rangle =\sqrt{\frac{1-\varepsilon}{2}}e^{i\varphi
_{\alpha_{1}}}\left\vert E_{1}\right\rangle +\sqrt{\frac{1+\varepsilon}{2}%
}e^{i\varphi_{\alpha_{2}}}\left\vert E_{2}\right\rangle \text{,} \label{c9}%
\end{equation}
and,%
\begin{equation}
\left\vert B\right\rangle =\sqrt{\frac{1-\varepsilon}{2}}e^{i\varphi
_{\beta_{1}}}\left\vert E_{1}\right\rangle +\sqrt{\frac{1+\varepsilon}{2}%
}e^{i\varphi_{\beta_{2}}}\left\vert E_{2}\right\rangle \text{,} \label{9b}%
\end{equation}
respectively. Therefore, employing Eqs. (\ref{c9}) and (\ref{9b}), the
orthogonality condition $\left\langle A\left\vert B\right.  \right\rangle =0$
reduces to%
\begin{equation}
\frac{1-\varepsilon}{2}e^{-i\left(  \varphi_{\alpha_{1}}-\varphi_{\beta_{1}%
}\right)  }+\frac{1+\varepsilon}{2}e^{-i\left(  \varphi_{\alpha_{2}}%
-\varphi_{\beta_{2}}\right)  }=0\text{.} \label{c10}%
\end{equation}
With no loss of generality, let us put%
\begin{equation}
e^{-i\left(  \varphi_{\alpha_{1}}-\varphi_{\beta_{1}}\right)  }=a+ib\text{,
and }e^{-i\left(  \varphi_{\alpha_{2}}-\varphi_{\beta_{2}}\right)
}=c+id\text{,} \label{c11}%
\end{equation}
with $a^{2}+b^{2}=1=c^{2}+d^{2}$, where $a$, $b$, $c$, and $d$ belong to $%
\mathbb{R}
$. Then, using Eq. (\ref{c11}), the complex constraint in Eq. (\ref{c10})
leads to the following two real constraints%
\begin{equation}
\left\{
\begin{array}
[c]{c}%
a(1-\varepsilon)+c\left(  1+\varepsilon\right)  =0\\
\left[  b\left(  1-\varepsilon\right)  +d\left(  1+\varepsilon\right)
\right]  =0
\end{array}
\right.  \text{.} \label{c12}%
\end{equation}
From Eq. (\ref{c12}), we get (being $\varepsilon\in\left(  0\text{, }1\right)
$)%
\begin{equation}
a=-c\frac{1+\varepsilon}{1-\varepsilon}\text{, and }b=-d\frac{1+\varepsilon
}{1-\varepsilon}\text{. } \label{c13}%
\end{equation}
Then, since we must have $a^{2}+b^{2}=1=c^{2}+d^{2}$ because of the
normalization conditions, Eq. (\ref{c13}) yields%
\begin{equation}
a^{2}+b^{2}=c^{2}\left(  \frac{1+\varepsilon}{1-\varepsilon}\right)
^{2}+d^{2}\left(  \frac{1+\varepsilon}{1-\varepsilon}\right)  ^{2}=\left(
c^{2}+d^{2}\right)  \left(  \frac{1+\varepsilon}{1-\varepsilon}\right)
^{2}=c^{2}+d^{2}\text{.} \label{c13b}%
\end{equation}
Therefore, Eq. (\ref{c13b}) is valid if and only if
\begin{equation}
\left(  \frac{1+\varepsilon}{1-\varepsilon}\right)  ^{2}=1\text{.} \label{c14}%
\end{equation}
Finally, we have that the condition in Eq. (\ref{c14}) is satisfied only if
$\varepsilon=0$ (which specifies the time-optimal scenario). Note that for
$\varepsilon=0$, the orthogonality condition in Eq. (\ref{c10}) is satisfied
by $\varphi_{\alpha_{1}}-\varphi_{\beta_{1}}=0$ and $\varphi_{\alpha_{2}%
}-\varphi_{\beta_{2}}=\pi$. In particular, denoting $\varphi_{\alpha_{2}%
}=\varphi_{\alpha}$ and $\varphi_{\beta_{2}}=\varphi_{\beta}$, the states
$\left\vert A\right\rangle $ and $\left\vert B\right\rangle $ in Eqs.
(\ref{c9}) and (\ref{9b}) assume the form%
\begin{equation}
\left\vert A\right\rangle =\frac{\left\vert E_{1}\right\rangle +e^{i\varphi
_{\alpha}}\left\vert E_{2}\right\rangle }{\sqrt{2}}\text{, and }\left\vert
B\right\rangle =\frac{\left\vert E_{1}\right\rangle +e^{i\varphi_{\beta}%
}\left\vert E_{2}\right\rangle }{\sqrt{2}}\text{,}%
\end{equation}
with $\left\langle A\left\vert B\right.  \right\rangle =e^{-i\frac
{\varphi_{\alpha}-\varphi_{\beta}}{2}}\cos\left[  \left(  \varphi_{\alpha
}-\varphi_{\beta}\right)  /2\right]  $ and, in addition,%
\begin{equation}
\left\vert \left\langle E_{1}\left\vert A\right.  \right\rangle \right\vert
^{2}=\frac{1}{2}=\left\vert \left\langle E_{2}\left\vert A\right.
\right\rangle \right\vert ^{2}\text{, and }\left\vert \left\langle
E_{1}\left\vert B\right.  \right\rangle \right\vert ^{2}=\frac{1}%
{2}=\left\vert \left\langle E_{2}\left\vert B\right.  \right\rangle
\right\vert ^{2}. \label{empty1}%
\end{equation}
Therefore, we recover the scenario that specifies the time-optimal evolution.
This concludes our energy-oriented explanation regarding the absence of paths
exceeding $\pi$ in length that link orthogonal states when their evolution is
restricted to the two-dimensional subspace they inhabit.

In summary, the sole method to facilitate suboptimal evolutions between
orthogonal states is to exit the two-dimensional motion or, alternatively, to
consider time-varying Hamiltonians to remain within two dimensions. In the
first scenario, one can create nongeodesic paths of length exceeding $\pi$
with constant Hamiltonians. Conversely, in the second scenario, nongeodesic
paths of length greater than $\pi$ can be examined because the unitary
evolution is defined by an initial state that intersects multiple projective
lines prior to reaching $\left\vert B\right\rangle $, with these
\textquotedblleft instantaneous\textquotedblright\ lines determined by a
time-dependent pair of energy eigenstates for the time-varying Hamiltonian
under consideration. We refer to Fig. $1$ for a sketch of the geometry of
Bloch vectors for both time optimal and time suboptimal evolutions between
orthogonal states specified by a constant and a nonstationary Hamiltonian, respectively.

The forthcoming section will focus on a detailed examination of the deviation
from time-optimality when utilizing a nonstationary Hamiltonian, all while
staying within a two-dimensional subspace.\begin{figure}[t]
\centering
\includegraphics[width=0.75\textwidth] {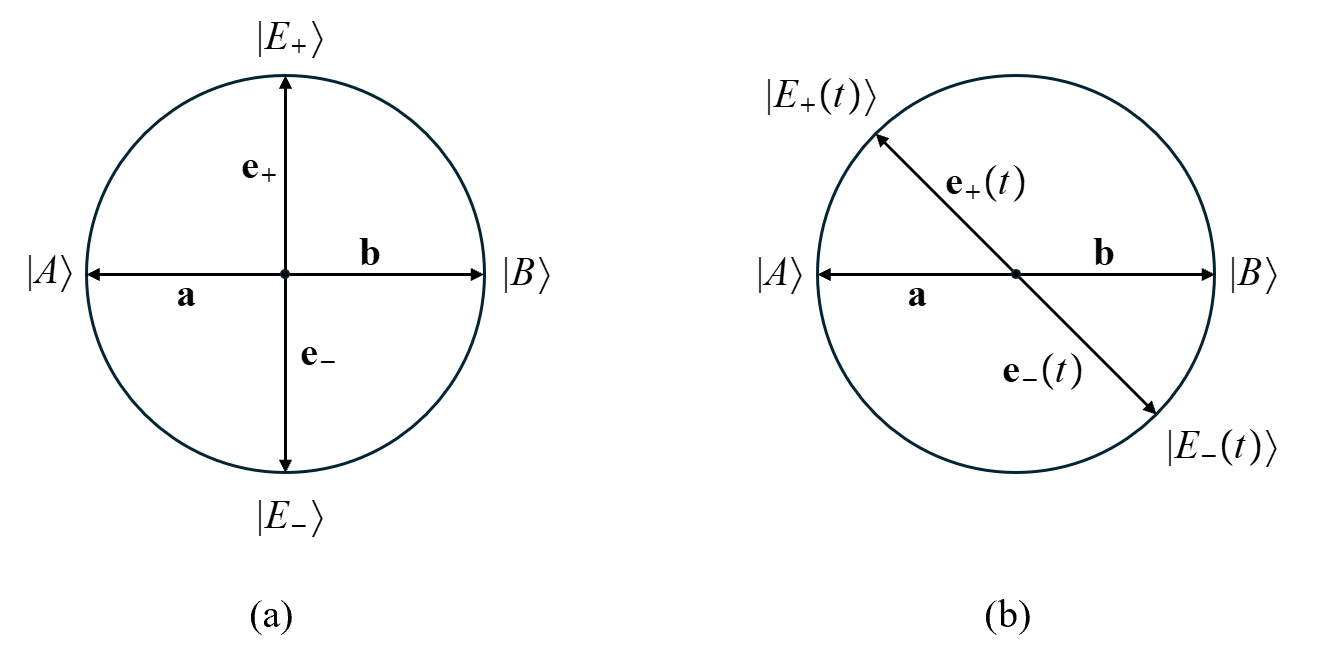}\caption{In (a), there is sketch
of the geometry of Bloch vectors for a time optimal evolution (with $\Delta
E=\Delta E_{\max}$) between orthogonal states $\left\vert A\right\rangle $ and
$\left\vert B\right\rangle $ specified by a constant Hamiltonian. In this
case, we have $\mathbf{a\cdot e}_{-}=0=\mathbf{a\cdot e}_{+}$ and
$\mathbf{b\cdot e}_{-}=0=\mathbf{b\cdot e}_{+}$. Clearly, $\mathbf{a\cdot
b=}-1\mathbf{=e}_{-}\cdot\mathbf{e}_{+}$. In (b), there is sketch of the
geometry of Bloch vectors for a time suboptimal evolution (with $\Delta
E\left(  t\right)  <\Delta E_{\max}$) between orthogonal states $\left\vert
A\right\rangle $ and $\left\vert B\right\rangle $ specified by a nonstationary
Hamiltonian. In this case, we generally have $\mathbf{a\cdot e}_{-}\left(
t\right)  \neq0=\mathbf{a\cdot e}_{+}\left(  t\right)  $ and $\mathbf{b\cdot
e}_{-}\left(  t\right)  =0=\mathbf{b\cdot e}_{+}\left(  t\right)  $. Clearly,
$\mathbf{a\cdot b=}-1\mathbf{=e}_{-}\left(  t\right)  \cdot\mathbf{e}%
_{+}\left(  t\right)  $.}%
\end{figure}

\section{Escaping time-optimality}

In this section, we provide an analytical demonstration of a distinct example
illustrating the possibility of deviating from the time-optimal evolution
between orthogonal states within a two-dimensional subspace of a larger
Hilbert space that they create, considering that the Hamiltonian may vary over time.

\subsection{Preliminaries}

In the subsequent discussion, we concentrate on single-qubit systems. In
accordance with Ref. \cite{uzdin12}, our objective is to identify an
appropriate time-dependent qubit Hamiltonian that propels the quantum system
with optimal speed efficiency along a non-geodesic trajectory on the Bloch
sphere (i.e., suboptimal geodesic efficiency). We note that the geodesic
efficiency of a quantum evolution from $\left\vert A\right\rangle $ and
$\left\vert B\right\rangle $ is defined as the ratio of the length of the
geodesic path that connects $\left\vert A\right\rangle $ and $\left\vert
B\right\rangle $ on the Bloch sphere to the length of the actual dynamical
path produced by the system's Hamiltonian
\cite{anandan90,cafaropra20,rossetti25}. Conversely, the speed efficiency of a
quantum evolution from $\left\vert A\right\rangle \overset{\text{def}}%
{=}\left\vert \psi\left(  0\right)  \right\rangle $ and $\left\vert
B\right\rangle \overset{\text{def}}{=}\left\vert \psi\left(  t_{\mathrm{final}%
}\right)  \right\rangle $ is typically an instantaneous time-dependent
quantity, where $0\leq t\leq t_{\mathrm{final}}$, and is characterized by the
ratio of the energy uncertainty of the system (i.e., the energy expended
during the quantum evolution) to the spectral norm of the Hamiltonian (i.e.,
the total energy of the system) \cite{uzdin12}.

In Ref. \cite{uzdin12}, the most comprehensive Hermitian nonstationary qubit
Hamiltonian $\mathrm{H}\left(  t\right)  $ is formulated in a way that it
produces the identical motion $\pi\left(  \left\vert \psi\left(  t\right)
\right\rangle \right)  $ within the complex projective Hilbert space $%
\mathbb{C}
P^{1}$ (or, correspondingly, on the Bloch sphere $S^{2}\cong%
\mathbb{C}
P^{1}$) as $\left\vert \psi\left(  t\right)  \right\rangle $, where the
projection operator $\pi$ is defined such that $\pi:\mathcal{H}_{2}^{1}%
\ni\left\vert \psi\left(  t\right)  \right\rangle \mapsto\pi\left(  \left\vert
\psi\left(  t\right)  \right\rangle \right)  \in%
\mathbb{C}
P^{1}$. Generally, it can be demonstrated that $\mathrm{H}\left(  t\right)  $
can be expressed as%
\begin{equation}
\mathrm{H}\left(  t\right)  =i\hslash\left\vert \partial_{t}m(t)\right\rangle
\left\langle m(t)\right\vert -i\hslash\left\vert m(t)\right\rangle
\left\langle \partial_{t}m(t)\right\vert \text{,} \label{optH}%
\end{equation}
where, for the sake of simplicity in notation, we can define $\left\vert
m(t)\right\rangle =\left\vert m\right\rangle $, $\left\vert \partial
_{t}m(t)\right\rangle =\left\vert \dot{m}\right\rangle $, and $\hslash=1$. The
unit state vector $\left\vert m\right\rangle $ satisfies the relations
$\pi\left(  \left\vert m(t)\right\rangle \right)  =\pi\left(  \left\vert
\psi\left(  t\right)  \right\rangle \right)  $ and $i\partial_{t}\left\vert
m(t)\right\rangle =\mathrm{H}\left(  t\right)  \left\vert m(t)\right\rangle $.
The condition $\pi\left(  \left\vert m(t)\right\rangle \right)  =\pi\left(
\left\vert \psi\left(  t\right)  \right\rangle \right)  $ leads to the
conclusion that $\left\vert m(t)\right\rangle =c(t)\left\vert \psi\left(
t\right)  \right\rangle $, with $c(t)$ denoting a complex function. By
stipulating that $\left\langle m\left\vert m\right.  \right\rangle =1$, we
conclude that $\left\vert c(t)\right\vert =1$. This subsequent condition
indicates, in turn, that $c(t)=e^{i\phi\left(  t\right)  }$ for a certain real
phase $\phi\left(  t\right)  $. Subsequently, by applying the parallel
transport condition $\left\langle m\left\vert \dot{m}\right.  \right\rangle
=\left\langle \dot{m}\left\vert m\right.  \right\rangle =0$, the phase
$\phi\left(  t\right)  $ is determined to be equal to $i\int\left\langle
\psi\left\vert \dot{\psi}\right.  \right\rangle dt$. Consequently, $\left\vert
m(t)\right\rangle =\exp(-\int_{0}^{t}\left\langle \psi(t^{\prime})\left\vert
\partial_{t^{\prime}}\psi(t^{\prime})\right.  \right\rangle dt^{\prime
})\left\vert \psi\left(  t\right)  \right\rangle $. It is important to note
that $\mathrm{H}\left(  t\right)  $ in Eq. (\ref{optH}) is inherently
traceless, as it contains only off-diagonal elements within the orthogonal
basis $\left\{  \left\vert m\right\rangle \text{, }\left\vert \partial
_{t}m\right\rangle \right\}  $. Moreover, the condition $i\partial
_{t}\left\vert m(t)\right\rangle =\mathrm{H}\left(  t\right)  \left\vert
m(t)\right\rangle $ indicates that $\left\vert m(t)\right\rangle $ adheres to
the Schr\"{o}dinger evolution equation. After providing some fundamental
preliminary information regarding Uzdin's research in Ref. \cite{uzdin12}, we
are now able to present our suggested time-dependent Hamiltonian.

\subsection{Hamiltonian model}

We start by examining a unit quantum state represented as%
\begin{equation}
\left\vert \psi\left(  t\right)  \right\rangle =\cos\left(  \omega
_{0}t\right)  \left\vert 0\right\rangle +e^{i\nu_{0}t}\sin\left(  \omega
_{0}t\right)  \left\vert 1\right\rangle \text{,}%
\end{equation}
where $\nu_{0}$ and $\omega_{0}$ are the two crucial real positive parameters
that can be adjusted while investigating the quantum evolution. Recall that,
using the polar and azimuthal angles $\theta\left(  t\right)  $ and
$\varphi\left(  t\right)  $, respectively, an arbitrary state on the Bloch
sphere can be expressed as $\left\vert \psi\left(  t\right)  \right\rangle
=\left\vert \psi\left(  \theta\left(  t\right)  \text{, }\varphi\left(
t\right)  \right)  \right\rangle =\cos\left[  \theta\left(  t\right)
/2\right]  \left\vert 0\right\rangle +e^{i\varphi\left(  t\right)  }%
\sin\left[  \theta\left(  t\right)  /2\right]  \left\vert 1\right\rangle $.
Consequently, we have $\omega_{0}t=\theta\left(  t\right)  /2$ and $\nu
_{0}t=\varphi\left(  t\right)  $, which means $\omega_{0}=\dot{\theta}/2$ and
$\nu_{0}=\dot{\varphi}$. Given that $\left\langle \psi\left(  t\right)
\left\vert \dot{\psi}\left(  t\right)  \right.  \right\rangle =i\nu_{0}%
\sin^{2}\left(  \omega_{0}t\right)  \neq0$, it follows that $\left\vert
\psi\left(  t\right)  \right\rangle $ is not parallel transported. Starting
from the state $\left\vert \psi\left(  t\right)  \right\rangle $, we define
the state $\left\vert m(t)\right\rangle \overset{\text{def}}{=}e^{-i\phi
\left(  t\right)  }\left\vert \psi\left(  t\right)  \right\rangle $ with the
phase $\phi\left(  t\right)  $ such that $\left\langle m(t)\left\vert \dot
{m}(t)\right.  \right\rangle =0$. A straightforward calculation shows that
$\left\langle m(t)\left\vert \dot{m}(t)\right.  \right\rangle =0$ if and only
if $-i\dot{\phi}+\left\langle \psi\left(  t\right)  \left\vert \dot{\psi
}\left(  t\right)  \right.  \right\rangle =0$, which implies $\dot{\phi
}=-i\left\langle \psi\left(  t\right)  \left\vert \dot{\psi}\left(  t\right)
\right.  \right\rangle $. By setting $\phi\left(  0\right)  =0$, the temporal
evolution of the phase $\phi\left(  t\right)  $ is described by%
\begin{equation}
\phi\left(  t\right)  =\frac{\nu_{0}}{4\omega_{0}}\left[  2\omega_{0}%
t-\sin\left(  2\omega_{0}t\right)  \right]  \text{.} \label{s5}%
\end{equation}
Moreover, by utilizing Eq. (\ref{s5}), the parallel transported unit state
$\left\vert m(t)\right\rangle $ is transformed into%
\begin{equation}
\left\vert m(t)\right\rangle =e^{-i\phi\left(  t\right)  }\left\vert
\psi\left(  t\right)  \right\rangle =e^{-i\frac{\nu_{0}}{4\omega_{0}}\left[
2\omega_{0}t-\sin\left(  2\omega_{0}t\right)  \right]  }\left[  \cos\left(
\omega_{0}t\right)  \left\vert 0\right\rangle +e^{i\nu_{0}t}\sin\left(
\omega_{0}t\right)  \left\vert 1\right\rangle \right]  \text{.} \label{s5a}%
\end{equation}
Note that $\left\vert A\right\rangle =\left\vert m(0)\right\rangle =\left\vert
0\right\rangle $, whereas for $t_{\ast}=\pi/(2\omega_{0})$ we find $\left\vert
B\right\rangle =\left\vert m\left(  t_{\ast}\right)  \right\rangle
\simeq\left\vert 1\right\rangle $ (with \textquotedblleft$\simeq
$\textquotedblright\ denoting physical equivalence of quantum states, modulo
unimportant global phase factors), leading to $\left\langle A\left\vert
B\right.  \right\rangle =\left\langle B\left\vert A\right.  \right\rangle =0$.
Consequently, we equate the matrix representation of the Hamiltonian
\textrm{H}$(t)=i(\left\vert \dot{m}\right\rangle \left\langle m\right\vert
-\left\vert m\right\rangle \left\langle \dot{m}\right\vert )$ in the
orthogonal basis $\left\{  \left\vert m\right\rangle \text{, }\left\vert
\dot{m}\right\rangle \right\}  $ to $\mathrm{H}(t)=h_{0}\left(  t\right)
\mathrm{I}+\mathbf{h}\left(  t\right)  \cdot\mathbf{\boldsymbol{\sigma}}$,
where $\left\vert m\right\rangle $ is defined in Eq. (\ref{s5a}). Furthermore,
by expressing $\rho\left(  t\right)  =\left\vert m(t)\right\rangle
\left\langle m(t)\right\vert =(1/2)\left[  \mathrm{I}+\mathbf{a}%
(t)\cdot\mathbf{\boldsymbol{\sigma}}\right]  $, we derive that the Bloch
vector $\mathbf{a}(t)$ and\textbf{ }the field vector $\mathbf{h}\left(
t\right)  $ become%
\begin{equation}
\mathbf{a}(t)=\left(
\begin{array}
[c]{c}%
\sin\left(  2\omega_{0}t\right)  \cos\left(  \nu_{0}t\right) \\
\sin\left(  \nu_{0}t\right)  \sin\left(  2\omega_{0}t\right) \\
\cos\left(  2\omega_{0}t\right)
\end{array}
\right)  \text{,} \label{BV}%
\end{equation}
and\textbf{,}%
\begin{equation}
\mathbf{h}\left(  t\right)  =\left(
\begin{array}
[c]{c}%
-\frac{\nu_{0}}{2}\cos\left(  2\omega_{0}t\right)  \sin(2\omega_{0}%
t)\cos\left(  \nu_{0}t\right)  -\omega_{0}\sin\left(  \nu_{0}t\right) \\
-\frac{\nu_{0}}{2}\cos\left(  2\omega_{0}t\right)  \sin(2\omega_{0}%
t)\sin\left(  \nu_{0}t\right)  +\omega_{0}\cos\left(  \nu_{0}t\right) \\
\frac{\nu_{0}}{2}\sin^{2}\left(  2\omega_{0}t\right)
\end{array}
\right)  \text{,} \label{m2}%
\end{equation}
respectively, with $h_{0}\left(  t\right)  =0$. It is noteworthy to mention
that the\textbf{ }two real-valued three-dimensional vectors $\mathbf{a}(t)$
and $\mathbf{h}\left(  t\right)  $ presented in Eqs. (\ref{BV}) and
(\ref{m2}), respectively, satisfy the vector differential equation
$\mathbf{\dot{a}}(t)=2\mathbf{h}\left(  t\right)  \times\mathbf{a}(t)$
\cite{feynman57,cafaropra25}. Indeed, up to constant factors, this vector
differential equation describes the Larmor precession. This phenomenon is
specified by the relation\textbf{ }$d\mathbf{m/}dt=\gamma\mathbf{m\times B}$,
where $\mathbf{m}$ is the magnetic moment vector, $\mathbf{B}$ denotes the
external\textbf{ }magnetic moment, and $\gamma$ is the gyromagnetic ratio.

Although the Hamiltonian models considered here are constructed from the
quantum state $\left\vert \psi\left(  t\right)  \right\rangle =\cos\left[
\alpha\left(  t\right)  \right]  \left\vert 0\right\rangle +e^{i\beta\left(
t\right)  }\sin\left[  \alpha\left(  t\right)  \right]  \left\vert
1\right\rangle $ with $\alpha\left(  t\right)  \overset{\text{def}}{=}%
\omega_{0}t$ and $\beta\left(  t\right)  \overset{\text{def}}{=}\nu_{0}t$, our
construction does not rely on this specific parametrization. It can be readily
shown that the approach remains valid for more general parametrizations, where
the parameters $\alpha\left(  t\right)  $ and $\beta\left(  t\right)  $ are
not required to vary linearly in time. The particular choice adopted here was
made solely for clarity of presentation.

\subsection{Evolution of probabilities}

Consider the Hamiltonian given by $\mathrm{H}\left(  t\right)  =i\left\vert
\dot{m}\right\rangle \left\langle m\right\vert -i\left\vert m\right\rangle
\left\langle \dot{m}\right\vert $, where $\left\langle m\left\vert m\right.
\right\rangle =1$, $\left\langle m\left\vert \dot{m}\right.  \right\rangle
=\left\langle \dot{m}\left\vert m\right.  \right\rangle =0$, and $\left\langle
\dot{m}\left\vert \dot{m}\right.  \right\rangle \neq1$. With respect to the
orthonormal basis $\mathcal{B}\overset{\text{def}}{\mathcal{=}}\left\{
\left\vert m\right\rangle \text{, }\left\vert \dot{m}\right\rangle
/\sqrt{\left\langle \dot{m}\left\vert \dot{m}\right.  \right\rangle }\right\}
$, we have%
\begin{equation}
\left[  \mathrm{H}\right]  _{\mathcal{B}}=\left(
\begin{array}
[c]{cc}%
\left\langle m\left\vert \mathrm{H}\right\vert m\right\rangle  &
\frac{\left\langle m\left\vert \mathrm{H}\right\vert \dot{m}\right\rangle
}{\sqrt{\left\langle \dot{m}\left\vert \dot{m}\right.  \right\rangle }}\\
\frac{\left\langle \dot{m}\left\vert \mathrm{H}\right\vert m\right\rangle
}{\sqrt{\left\langle \dot{m}\left\vert \dot{m}\right.  \right\rangle }} &
\frac{\left\langle \dot{m}\left\vert \mathrm{H}\right\vert \dot{m}%
\right\rangle }{\left\langle \dot{m}\left\vert \dot{m}\right.  \right\rangle }%
\end{array}
\right)  =\left(
\begin{array}
[c]{cc}%
0 & -i\sqrt{\left\langle \dot{m}\left\vert \dot{m}\right.  \right\rangle }\\
i\sqrt{\left\langle \dot{m}\left\vert \dot{m}\right.  \right\rangle } & 0
\end{array}
\right)  \text{.} \label{face2}%
\end{equation}
Searching for eigenvalues and eigenvectors of $\left[  \mathrm{H}\right]
_{\mathcal{B}}$ in Eq. (\ref{face2}), we find that they are given by%
\begin{equation}
E_{+}\left(  t\right)  \overset{\text{def}}{=}+\sqrt{\left\langle \dot
{m}\left\vert \dot{m}\right.  \right\rangle }\text{, }E_{-}\left(  t\right)
\overset{\text{def}}{=}-\sqrt{\left\langle \dot{m}\left\vert \dot{m}\right.
\right\rangle }\text{,}%
\end{equation}
and%
\begin{equation}
\left\vert E_{+}\left(  t\right)  \right\rangle \overset{\text{def}}{=}%
\frac{1}{\sqrt{2}}\left[  \left\vert m\right\rangle +i\frac{\left\vert \dot
{m}\right\rangle }{\sqrt{\left\langle \dot{m}\left\vert \dot{m}\right.
\right\rangle }}\right]  \text{, }\left\vert E_{-}\left(  t\right)
\right\rangle \overset{\text{def}}{=}\frac{1}{\sqrt{2}}\left[  \left\vert
m\right\rangle -i\frac{\left\vert \dot{m}\right\rangle }{\sqrt{\left\langle
\dot{m}\left\vert \dot{m}\right.  \right\rangle }}\right]  \text{,}
\label{cafe111}%
\end{equation}
respectively. Note that, for the sake of convenience in notations, we have
substituted $\left\{  \left\vert E_{1}\right\rangle \text{, }\left\vert
E_{2}\right\rangle \right\}  $ with $\left\{  \left\vert E_{+}\right\rangle
\text{, }\left\vert E_{-}\right\rangle \right\}  $. Therefore, the spectral
decomposition of $\mathrm{H}=i\left\vert \dot{m}\right\rangle \left\langle
m\right\vert -i\left\vert m\right\rangle \left\langle \dot{m}\right\vert $ is
given by%
\begin{equation}
\mathrm{H}\left(  t\right)  =E_{+}\left(  t\right)  \left\vert E_{+}\left(
t\right)  \right\rangle \left\langle E_{+}\left(  t\right)  \right\vert
+E_{-}\left(  t\right)  \left\vert E_{-}\left(  t\right)  \right\rangle
\left\langle E_{-}\left(  t\right)  \right\vert \text{.} \label{face3}%
\end{equation}

\begin{figure}[t]
\centering
\includegraphics[width=1\textwidth] {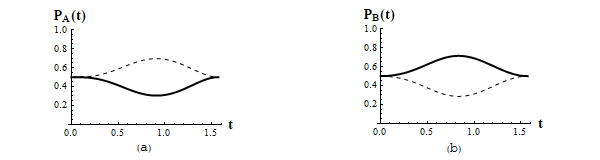}\caption{In (a), there are the plots
of the probabilities \textrm{P}$_{A}\left(  t\right)  \overset{\text{def}}%
{=}\left\vert \left\langle E_{+}(t)\left\vert A\right.  \right\rangle
\right\vert ^{2}$ (dashed line) and \textrm{P}$_{A}\left(  t\right)
\overset{\text{def}}{=}\left\vert \left\langle E_{-}(t)\left\vert A\right.
\right\rangle \right\vert ^{2}$ (thick solid line) versus time $t$. In (b),
there are the plots of the probabilities \textrm{P}$_{B}\left(  t\right)
\overset{\text{def}}{=}\left\vert \left\langle E_{+}(t)\left\vert B\right.
\right\rangle \right\vert ^{2}$ (dashed line) and \textrm{P}$_{B}\left(
t\right)  \overset{\text{def}}{=}\left\vert \left\langle E_{-}(t)\left\vert
B\right.  \right\rangle \right\vert ^{2}$ (thick solid line) versus time $t$.
In all plots, we set $\nu_{0}=\omega_{0}=1$ and $0\leq t\leq\pi/2$.}%
\end{figure}

It is a straightforward to verify that $\mathrm{H}\left(  t\right)
=i\left\vert \dot{m}\right\rangle \left\langle m\right\vert -i\left\vert
m\right\rangle \left\langle \dot{m}\right\vert $ coincides with $\mathrm{H}%
\left(  t\right)  $ in Eq. (\ref{face3}). Now, considering an evolution from
$\left\vert A\right\rangle $ to $\left\vert B\right\rangle $ with
$\left\langle A\left\vert B\right.  \right\rangle =\left\langle B\left\vert
A\right.  \right\rangle =0$ (and, additionally, assuming both states
normalized to one), we have%
\begin{equation}
\left\vert A\right\rangle =\left\langle E_{+}\left(  t\right)  \left\vert
A\right.  \right\rangle \left\vert E_{+}\left(  t\right)  \right\rangle
+\left\langle E_{-}\left(  t\right)  \left\vert A\right.  \right\rangle
\left\vert E_{-}\left(  t\right)  \right\rangle \text{, and }\left\vert
B\right\rangle =\left\langle E_{+}\left(  t\right)  \left\vert B\right.
\right\rangle \left\vert E_{+}\left(  t\right)  \right\rangle +\left\langle
E_{-}\left(  t\right)  \left\vert B\right.  \right\rangle \left\vert
E_{-}\left(  t\right)  \right\rangle \text{.} \label{cafe2}%
\end{equation}
Using Eqs. (\ref{cafe111}) and (\ref{cafe2}), we notice that we generally
obtain%
\begin{equation}
\left\vert \left\langle E_{+}\left(  t\right)  \left\vert A\right.
\right\rangle \right\vert ^{2}=\frac{1}{2}\left\vert \left\langle m\left\vert
A\right.  \right\rangle -i\frac{\left\langle \dot{m}\left\vert A\right.
\right\rangle }{\sqrt{\left\langle \dot{m}\left\vert \dot{m}\right.
\right\rangle }}\right\vert ^{2}\neq\frac{1}{2}\left\vert \left\langle
m\left\vert A\right.  \right\rangle +i\frac{\left\langle \dot{m}\left\vert
A\right.  \right\rangle }{\sqrt{\left\langle \dot{m}\left\vert \dot{m}\right.
\right\rangle }}\right\vert ^{2}=\left\vert \left\langle E_{-}\left(
t\right)  \left\vert A\right.  \right\rangle \right\vert ^{2}\text{,}
\label{cafe3}%
\end{equation}
with $\left\vert \left\langle E_{+}\left(  t\right)  \left\vert A\right.
\right\rangle \right\vert ^{2}+\left\vert \left\langle E_{-}\left(  t\right)
\left\vert A\right.  \right\rangle \right\vert ^{2}=1$. More explicitly,
taking $\left\vert A\right\rangle =\left\vert 0\right\rangle $ and $\left\vert
m\right\rangle =e^{-i\phi}\left[  \cos(\alpha)\left\vert 0\right\rangle
+e^{i\beta}\sin(\alpha)\left\vert 1\right\rangle \right]  $ where
$\alpha\left(  t\right)  \overset{\text{def}}{=}\omega_{0}t$ and
$\beta(t)\overset{\text{def}}{=}\nu_{0}t$, the probability amplitude squared
$\left\vert \left\langle E_{+}\left(  t\right)  \left\vert A\right.
\right\rangle \right\vert ^{2}$ in Eq. (\ref{cafe3}) reduces to%
\begin{equation}
\left\vert \left\langle E_{+}\left(  t\right)  \left\vert A\right.
\right\rangle \right\vert ^{2}=\frac{1}{2}\left\vert \left(  \cos
(\alpha)+\frac{\dot{\phi}\cos(\alpha)}{\sqrt{\dot{\alpha}^{2}+\frac{1}{4}%
\dot{\beta}^{2}\sin^{2}(2\alpha)}}\right)  +i\frac{\dot{\alpha}\sin(\alpha
)}{\sqrt{\dot{\alpha}^{2}+\frac{1}{4}\dot{\beta}^{2}\sin^{2}(2\alpha)}%
}\right\vert ^{2}\text{,} \label{face4}%
\end{equation}
while $\left\vert \left\langle E_{-}\left(  t\right)  \left\vert A\right.
\right\rangle \right\vert ^{2}$ in Eq. (\ref{cafe3}) becomes,%
\begin{equation}
\left\vert \left\langle E_{-}\left(  t\right)  \left\vert A\right.
\right\rangle \right\vert ^{2}=\frac{1}{2}\left\vert \left(  \cos
(\alpha)-\frac{\dot{\phi}\cos(\alpha)}{\sqrt{\dot{\alpha}^{2}+\frac{1}{4}%
\dot{\beta}^{2}\sin^{2}(2\alpha)}}\right)  -i\frac{\dot{\alpha}\sin(\alpha
)}{\sqrt{\dot{\alpha}^{2}+\frac{1}{4}\dot{\beta}^{2}\sin^{2}(2\alpha)}%
}\right\vert ^{2}\text{.} \label{face5}%
\end{equation}
As a side remark, one can easily verify by means of Eqs. (\ref{face4}) and
(\ref{face5}) that $\left\vert \left\langle E_{+}\left(  t\right)  \left\vert
A\right.  \right\rangle \right\vert ^{2}+\left\vert \left\langle E_{-}\left(
t\right)  \left\vert A\right.  \right\rangle \right\vert ^{2}=1$ using some
trigonometry and the fact that $\dot{\phi}=\dot{\beta}\sin^{2}(\alpha)$.
Similarly, taking $\left\vert B\right\rangle =\left\vert 1\right\rangle $ and
$\left\vert m\right\rangle =e^{-i\phi}\left[  \cos(\alpha)\left\vert
0\right\rangle +e^{i\beta}\sin(\alpha)\left\vert 1\right\rangle \right]  $,
use of Eqs. (\ref{cafe111}) and (\ref{cafe2}) yields probability amplitude
squared $\left\vert \left\langle E_{+}\left(  t\right)  \left\vert B\right.
\right\rangle \right\vert ^{2}$ and $\left\vert \left\langle E_{-}\left(
t\right)  \left\vert B\right.  \right\rangle \right\vert ^{2}$ given by
\begin{equation}
\left\vert \left\langle E_{+}\left(  t\right)  \left\vert B\right.
\right\rangle \right\vert ^{2}=\frac{1}{2}\left\vert \left(  \sin
(\alpha)+\frac{\dot{\phi}\sin(\alpha)-\dot{\beta}\sin(\alpha)}{\sqrt
{\dot{\alpha}^{2}+\frac{1}{4}\dot{\beta}^{2}\sin^{2}(2\alpha)}}\right)
-i\frac{\dot{\alpha}\cos(\alpha)}{\sqrt{\dot{\alpha}^{2}+\frac{1}{4}\dot
{\beta}^{2}\sin^{2}(2\alpha)}}\right\vert ^{2}\text{,} \label{face4b}%
\end{equation}
and,%
\begin{equation}
\left\vert \left\langle E_{-}\left(  t\right)  \left\vert B\right.
\right\rangle \right\vert ^{2}=\frac{1}{2}\left\vert \left(  \sin
(\alpha)-\frac{\dot{\phi}\sin(\alpha)-\dot{\beta}\sin(\alpha)}{\sqrt
{\dot{\alpha}^{2}+\frac{1}{4}\dot{\beta}^{2}\sin^{2}(2\alpha)}}\right)
+i\frac{\dot{\alpha}\cos(\alpha)}{\sqrt{\dot{\alpha}^{2}+\frac{1}{4}\dot
{\beta}^{2}\sin^{2}(2\alpha)}}\right\vert ^{2}\text{,} \label{face5b}%
\end{equation}
respectively. Again, one can check that $\left\vert \left\langle E_{+}\left(
t\right)  \left\vert B\right.  \right\rangle \right\vert ^{2}+\left\vert
\left\langle E_{-}\left(  t\right)  \left\vert B\right.  \right\rangle
\right\vert ^{2}=1$ using some trigonometry and the fact that $\dot{\phi}%
=\dot{\beta}\sin^{2}(\alpha)$. For a plot of the transition probabilities
in\ Eqs. (\ref{face4}), (\ref{face5}), (\ref{face4b}), and (\ref{face5b}), we
refer to Fig. $2$. Moreover, if we set $\rho_{A}\overset{\text{def}}%
{=}\left\vert A\right\rangle \left\langle A\right\vert =(\mathbf{1+a}%
\cdot\mathbf{\boldsymbol{\sigma}})/2$, $\rho_{B}\overset{\text{def}}%
{=}\left\vert B\right\rangle \left\langle B\right\vert =(\mathbf{1+b}%
\cdot\mathbf{\boldsymbol{\sigma}})/2$, $\rho_{E_{\pm}}\overset{\text{def}}%
{=}\left\vert E_{\pm}\right\rangle \left\langle E_{\pm}\right\vert
=(\mathbf{1+e}_{\pm}\cdot\mathbf{\boldsymbol{\sigma}})/2$, and $\rho\left(
t\right)  \overset{\text{def}}{=}\left\vert \psi\left(  t\right)
\right\rangle \left\langle \psi\left(  t\right)  \right\vert =(\mathbf{1+a}%
\left(  t\right)  \cdot\mathbf{\boldsymbol{\sigma}})/2$, we note that
$\left\vert \left\langle A\left\vert B\right.  \right\rangle \right\vert
^{2}=\left(  1+\mathbf{a}\cdot\mathbf{b}\right)  /2$ since $\left\vert
\left\langle A\left\vert B\right.  \right\rangle \right\vert ^{2}%
=\mathrm{tr}\left(  \rho_{A}\rho_{B}\right)  +2\sqrt{\det\left(  \rho
_{A}\right)  \det\left(  \rho_{B}\right)  }$. Therefore, in the stationary
optimal-time evolution between orthogonal states $\left\vert A\right\rangle $
and $\left\vert B\right\rangle $, we have $\Delta E=\Delta E_{\max}$,
$\left\vert \left\langle E_{+}\left\vert A\right.  \right\rangle \right\vert
^{2}=1/2=\left\vert \left\langle E_{-}\left\vert A\right.  \right\rangle
\right\vert ^{2}$, $\mathbf{a}\cdot\mathbf{e}_{+}=0=\mathbf{a}\cdot
\mathbf{e}_{-}$, and the path length of the actual dynamical trajectory equals
the length of the geodesic path (i.e., $s=s_{\mathrm{geo}}=\theta_{AB}%
\overset{\text{def}}{=}2\arccos\left[  \left\vert \left\langle A\left\vert
B\right.  \right\rangle \right\vert \right]  =\pi$, where $s\overset
{\text{def}}{=}(2/\hslash)\int_{t_{A}}^{t_{B}}\Delta E\left(  t^{\prime
}\right)  dt^{\prime}$ \cite{anandan90}). Similar conditions in terms of
probability amplitudes and Bloch vectors hold true for the final state
$\left\vert B\right\rangle $ as well. Moreover, given that the Hamiltonian
\textrm{H} remains constant, similar conditions are also fulfilled by
$\left\vert \psi\left(  t\right)  \right\rangle $. From a geometric
standpoint, the key aspect to highlight is the orthogonality between the Bloch
vectors $\left\{  \mathbf{a}\text{, }\mathbf{b}\right\}  $ and $\left\{
\mathbf{e}_{+}\text{, }\mathbf{e}_{-}\right\}  $ throughout the evolution. Of
course, since $\left\vert A\right\rangle $ and $\left\vert B\right\rangle $
are orthogonal, we have that $\mathbf{a}$ and $\mathbf{b}$ are antiparallel
(i.e., $\mathbf{a\cdot b=}-1$). In the nonstationary time suboptimal
evolution, instead, we typically have $\Delta E=\Delta E\left(  t\right)
<\Delta E_{\max}$, $\left\vert \left\langle E_{+}\left(  t\right)  \left\vert
A\right.  \right\rangle \right\vert ^{2}\neq1/2\neq\left\vert \left\langle
E_{-}\left(  t\right)  \left\vert A\right.  \right\rangle \right\vert ^{2}$,
$\mathbf{a}\cdot\mathbf{e}_{+}\left(  t\right)  \neq0\neq\mathbf{a}%
\cdot\mathbf{e}_{-}\left(  t\right)  $, and the path length of the actual
dynamical trajectory is longer than the length of the geodesic path (i.e.,
$s>s_{\mathrm{geo}}$). From a geometric standpoint, there is typically an
absence of orthogonality between the Bloch vectors $\left\{  \mathbf{a}\text{,
}\mathbf{b}\right\}  $ and the Bloch vectors $\left\{  \mathbf{e}_{+}\text{,
}\mathbf{e}_{-}\right\}  $ throughout the quantum evolution. In fact, the
generation of time suboptimal evolutions between orthogonal states in a
two-dimensional space is quite feasible with time-dependent Hamiltonians, as
it allows for a deviation from the orthogonality between $\left\{
\mathbf{a}\text{, }\mathbf{b}\right\}  $ and $\left\{  \mathbf{e}_{+}\text{,
}\mathbf{e}_{-}\right\}  $. In conclusion, the symmetric arrangement that
presents itself through the local (i.e., instantaneous) orthogonality between
the Bloch vector $\mathbf{a}\left(  t\right)  $ with $0\leq t\leq
t_{\mathrm{final}}$ and the Bloch eigenvectors $\left\{  \mathbf{e}_{+}\text{,
}\mathbf{e}_{-}\right\}  $ prevents the emergence of dynamical paths with
lengths other than $\pi$ when evolving with a constant Hamiltonian between
orthogonal states within a two-dimensional subspace of the larger Hilbert
space. Considerations regarding optimal stationary and suboptimal
nonstationary Hamiltonian evolutions, focusing on path lengths, energy
uncertainties, probability amplitudes, and configurations of the Bloch vector
appear in Table I.

We are now prepared to offer physical insights into how one can avert analog
quantum search failures by drawing comparisons with occurrences in standard
quantum evolutions between specified initial and final states.\begin{table}[t]
\centering
\begin{tabular}
[c]{c|c|c}\hline\hline
\textbf{Type of constraint} & \textbf{Optimal stationary evolution} &
\textbf{Suboptimal nonstationary evolution}\\\hline
Path length & $s=\pi$ & $s>\pi$\\\hline
Energy uncertainty & $\Delta E=\Delta E_{\max}$ & $\Delta E\left(  t\right)
<\Delta E_{\max}$\\\hline
Probability amplitudes & $\left\vert \left\langle E_{+}\left\vert A\right.
\right\rangle \right\vert ^{2}=\frac{1}{2}=\left\vert \left\langle
E_{-}\left\vert A\right.  \right\rangle \right\vert ^{2}$ & $\left\vert
\left\langle E_{+}\left(  t\right)  \left\vert A\right.  \right\rangle
\right\vert ^{2}\neq\frac{1}{2}\neq\left\vert \left\langle E_{-}t\left\vert
A\right.  \right\rangle \right\vert ^{2}$\\\hline
Bloch vectors & $\mathbf{a\cdot e}_{+}=0=\mathbf{a\cdot e}_{-}$ &
$\mathbf{a\cdot e}_{+}\left(  t\right)  \neq0\neq\mathbf{a\cdot e}_{-}\left(
t\right)  $\\\hline
\end{tabular}
\caption{Aspects of optimal stationary and suboptimal nonstationary
Hamiltonian evolutions in terms of path lengths, energy uncertainties,
probability amplitudes, and Bloch vector configurations.}%
\end{table}

\section{Escaping analog quantum search failures}

In this section, we explore the concept of symmetry to pinpoint the source of
failures in continuous time quantum search schemes involving orthogonal source
and target states. Additionally, we propose methods to eliminate these
failures and, crucially, we establish an analogy with the failures observed in
symmetric configurations that define time-optimal evolutions between
orthogonal states governed by a stationary Hamiltonian.

\subsection{Time-independent setting}

The optimal time evolution between known initial and final states $\left\vert
A\right\rangle $ and $\left\vert B\right\rangle $, respectively, is a distinct
problem from that addressed in continuous-time quantum searching. In the
latter scenario, the search is conducted from a known source state $\left\vert
s\right\rangle $ to an unknown target state $\left\vert w\right\rangle $.
Specifically, because the target state remains unknown, the modulus of the
quantum overlap $\left\vert \left\langle s\left\vert w\right.  \right\rangle
\right\vert $ between the source and target states is also indeterminate.
However, these two quantum-mechanical processes possess a shared key feature.
When both processes are defined by stationary Hamiltonians, and evolutions
take place within the two-dimensional subspaces formed by the orthogonal sets
of states $\left\{  \left\vert A\right\rangle \text{, }\left\vert
B\right\rangle \right\}  $ or $\left\{  \left\vert s\right\rangle \text{,
}\left\vert w\right\rangle \right\}  $, the inability to surpass the
time-optimality of quantum evolutions is linked to the shortcomings of analog
quantum search methods. More precisely, the inadequacy of \textrm{H}%
$_{\mathrm{FG}}$ in\ Eq. (\ref{FG}) is reflected in the asymptotically
extended duration of the search scheme, as indicated by $t_{\mathrm{FG}}$ in
Eq. (\ref{tgf}). Notably, the inadequacy associated with \textrm{H}%
$_{\mathrm{Fenner}}$ in Eq. (\ref{Fenner}) is not apparent, as we observe that
Fenner's Hamiltonian is constructed in a way that inherently prevents
orthogonality between the source and target states. Indeed, we can
substantiate this assertion by observing that \textrm{H}$_{\mathrm{Fenner}}$
in Eq. (\ref{Fenner}) represents a specific instance of \textrm{H}%
$_{\mathrm{opt}}$ in Eq. (\ref{amy}). To confirm this assertion, it suffices
to set $\left\vert \left\langle A\left\vert B\right.  \right\rangle
\right\vert =x$ and note that $\Delta E_{\mathrm{Fenner}}$ simplifies to
$\Delta E_{\mathrm{Fenner}}\left(  x\right)  \overset{\text{def}}{=}%
2xE\sqrt{1-x^{2}}$. Furthermore, while both \textrm{H}$_{\mathrm{FG}}$ and
\textrm{H}$_{\mathrm{Fenner}}$ are optimal from the perspective of quantum
search (in the sense that $t_{\mathrm{FG}}\approx t_{\mathrm{Fenner}}%
\propto\sqrt{N}$, with $\sqrt{N}$ being the defining characteristic of
Grover's algorithm quadratic speedup), only \textrm{H}$_{\mathrm{Fenner}}$ is
genuinely time-optimal, as evidenced by $t_{\mathrm{Fenner}}=t_{\mathrm{opt}%
}^{\mathrm{H}_{\mathrm{Fenner}}}(x$, $E)$ (in contrast, $t_{\mathrm{FG}%
}>t_{\mathrm{opt}}^{\mathrm{H}_{\mathrm{FG}}}(x$, $E)$).

\subsection{Time-dependent setting}

To avoid failures in continuous-time quantum search between orthogonal source
and target states, inspired by the approach of employing time-dependent
Hamiltonians to create suboptimal evolutions between orthogonal states in
two-dimensional subspaces, one might also explore time-dependent quantum
search Hamiltonians \cite{farhi00,roland02,romanelli07,carloijqi}. In this
context, it is particularly crucial to highlight the principles of quantum
search through local adiabatic \cite{roland02} (or, alternatively,
nonadiabatic \cite{romanelli07}) evolutions. Within the framework of local
adiabatic processes, it is observed that by adjusting the evolution rate of
the Hamiltonian to ensure adiabatic evolution throughout each infinitesimal
time interval, the total running time becomes proportional to $\sqrt{N}$.
Notably, proper working of adiabatic quantum search Hamiltonians necessitate a
nonvanishing minimum energy gap, which is defined as the smallest value of the
energy difference between the two lowest instantaneous energy levels of the
Hamiltonian, calculated over the temporal duration of the search. For
instance, based on the research conducted by Roland and Cerf in
Ref.\cite{roland02}, the local adiabatic search Hamiltonian is characterized
as
\begin{equation}
\mathrm{H}_{\mathrm{RC}}(t)\overset{\text{def}}{=}(1-\frac{t}{T}%
)\mathrm{H}_{s}+\frac{t}{T}\mathrm{H}_{w}=\mathrm{\tilde{H}}\left(
\xi\right)  =(1-\xi)\mathrm{H}_{s}+\xi\mathrm{H}_{w}\text{,} \label{HRC}%
\end{equation}
with $0\leq\xi\leq1$, $\mathrm{H}_{s}\overset{\text{def}}{=}\mathbf{1+}%
\left\vert s\right\rangle \left\langle s\right\vert $, $\mathrm{H}_{w}%
\overset{\text{def}}{=}\mathbf{1+}\left\vert w\right\rangle \left\langle
w\right\vert $ and, finally $\left\vert s\right\rangle =\left(  1/\sqrt
{N}\right)  \sum_{i}\left\vert i\right\rangle $ with $\left\langle i\left\vert
j\right.  \right\rangle =\delta_{ij}$ for any $1\leq i$, $j\leq N$. A
symmetric scenario occurs, for instance, when $\mathrm{H}_{s}$ and
$\mathrm{H}_{w}$ in the decomposition of $\mathrm{\tilde{H}}\left(
\xi\right)  $ are diagonal in the same basis. In this case, the energy levels
$E_{0}\left(  \xi\right)  $ and $E_{1}\left(  \xi\right)  $ (with
$E_{0}\left(  \xi\right)  \leq E_{1}\left(  \xi\right)  \leq...\leq
E_{N-1}(\xi)$) of the Hamiltonian cross and, therefore, the minimum gap
defined as $g_{\min}\overset{\text{def}}{=}\underset{0\leq\xi\leq1}{\min
}\left[  E_{1}\left(  \xi\right)  -E_{0}\left(  \xi\right)  \right]  $
vanishes \cite{farhi00}. This, in turn, leads to the failure of the searching
scheme. In Fig. $3$, we depict this fact. Interestingly, the issues with
orthogonal states can also propagate to searching schemes specified by
time-dependent Hamiltonians. For example, if one considers \textrm{\~{H}%
}$\left(  \xi\right)  $ in Eq. (\ref{HRC}), problems emerge when $\left\vert
s\right\rangle \overset{\text{def}}{=}\left\vert 0\right\rangle $ and
$\left\vert w\right\rangle \overset{\text{def}}{=}\left\vert 1\right\rangle
\in\left\{  \left\vert 0\right\rangle \text{, }\left\vert 1\right\rangle
\right\}  $. Indeed, in this case \textrm{\~{H}}$\left(  \xi\right)  $
commutes with $\sigma_{z}$ for any $s$ and there is a level crossing that
makes $g_{\min}$ equal to zero for some $\xi$. A similar situation arises when
$\left\vert s\right\rangle \overset{\text{def}}{=}\left[  \left\vert
0\right\rangle +\left\vert 1\right\rangle \right]  /\sqrt{2}$ and $\left\vert
w\right\rangle \overset{\text{def}}{=}\left[  \left\vert 0\right\rangle
-\left\vert 1\right\rangle \right]  /\sqrt{2}\notin\left\{  \left\vert
0\right\rangle \text{, }\left\vert 1\right\rangle \right\}  $. Indeed, in this
case \textrm{\~{H}}$\left(  \xi\right)  $ commutes with $\sigma_{x}$ for any
$s$ and there exists a level crossing that causes $g_{\min}$ to vanish for
some $\xi$. Therefore, because of the existence of a symmetry, the search
scheme fails to work. More broadly, it can be noted that $\mathrm{\tilde{H}%
}\left(  \xi\right)  $ in Eq. (\ref{HRC}) satisfies the condition $\left[
\mathrm{H}_{s}\text{, }\mathrm{H}_{w}\right]  =0$ when $\left\langle
s\left\vert w\right.  \right\rangle =\left\langle w\left\vert s\right.
\right\rangle =0$. Consequently, when the source and target states are
orthogonal, it follows that $\left[  \mathrm{\tilde{H}}\text{, }\mathrm{H}%
_{s}\right]  =\left[  \mathrm{\tilde{H}}\text{, }\mathrm{H}_{w}\right]  =0$.
Once more, symmetries may result in search failures.\begin{figure}[t]
\centering
\includegraphics[width=0.5\textwidth] {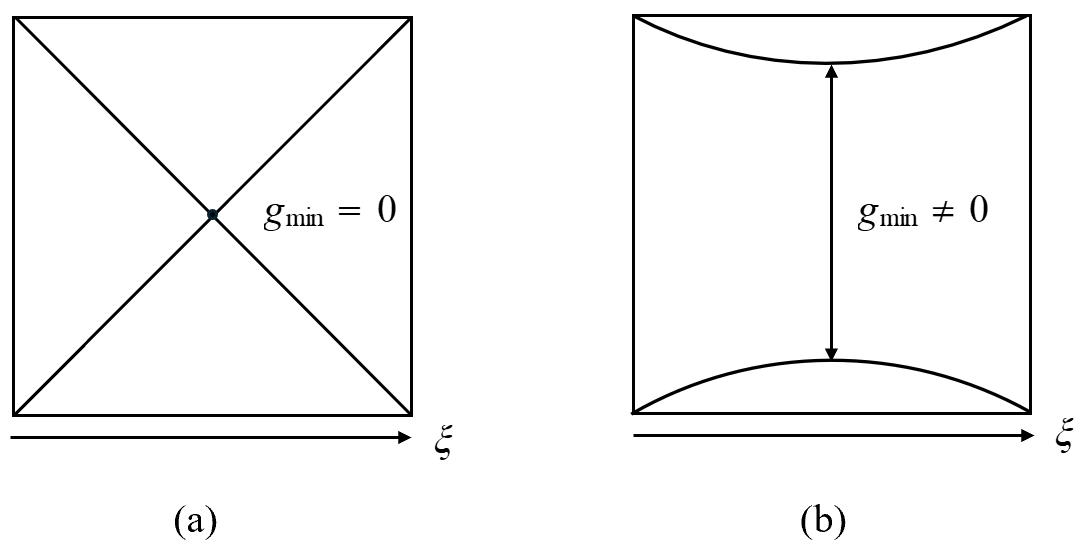}\caption{In (a), there is a sketch
of the two eigenvalues $\xi$ and $1-\xi$ of the time-dependent Hamiltonian
\textrm{\~{H}}$\left(  \xi\right)  \overset{\text{def}}{=}(1-\xi)\left[
\mathbf{1+}\left\vert s\right\rangle \left\langle s\right\vert \right]
+\xi\left[  \mathbf{1+}\left\vert w\right\rangle \left\langle w\right\vert
\right]  $ when $\left\vert s\right\rangle \overset{\text{def}}{=}\left\vert
0\right\rangle $, $\left\vert w\right\rangle \overset{\text{def}}{=}\left\vert
1\right\rangle $, $\left\vert \left\langle s\left\vert w\right.  \right\rangle
\right\vert =0$, and the reduced time $\xi$ being such that $0\leq\xi\leq1$.
In this case, $\mathbf{1+}\left\vert s\right\rangle \left\langle s\right\vert
$ and $\mathbf{1+}\left\vert w\right\rangle \left\langle w\right\vert $ are
diagonal in the same basis, the eigenvalues cross and, therefore, $g_{\min}%
=0$. In (b), there is a sketch of the two eigenvalues $(1+\sqrt{\left(
1-\xi\right)  ^{2}+\xi^{2}})/2$ and $(1-\sqrt{\left(  1-\xi\right)  ^{2}%
+\xi^{2}})/2$ of the time-dependent Hamiltonian \textrm{\~{H}}$\left(
\xi\right)  \overset{\text{def}}{=}(1-\xi)\left[  \mathbf{1+}\left\vert
s\right\rangle \left\langle s\right\vert \right]  +\xi\left[  \mathbf{1+}%
\left\vert w\right\rangle \left\langle w\right\vert \right]  $ when
$\left\vert s\right\rangle \overset{\text{def}}{=}\left\vert 0\right\rangle $,
$\left\vert w\right\rangle \overset{\text{def}}{=}\left[  \left\vert
0\right\rangle +\left\vert 1\right\rangle \right]  /2$, $\left\vert
\left\langle s\left\vert w\right.  \right\rangle \right\vert \neq0$, and
$0\leq\xi\leq1$. In this case, $\mathbf{1+}\left\vert s\right\rangle
\left\langle s\right\vert $ and $\mathbf{1+}\left\vert w\right\rangle
\left\langle w\right\vert $ are not diagonal in the same basis, the
eigenvalues do not cross and, therefore, the minimum energy gap does not
vanish since $g_{\min}\neq0$.}%
\end{figure}

What is the role played by symmetry in the failure of quantum searching
schemes? How is it related to the inability of transitioning between
orthogonal states with a constant Hamiltonian in a sub-optimal time fashion?
We recall that when the Hamiltonian of a system commutes with an operator,
there is a symmetry in the system. We have that: i) The dynamics of the system
is invariant under transformations generated by the operator; ii) The
observable corresponding to the non-explicitly time-dependent operator is
conserved in time; iii) The Hamiltonian and the operator share a common set of
eigenstates, so they can be simultaneously diagonalized. For example, if
$\left[  \mathrm{H}\text{, }\sigma_{z}\right]  =0$, the system is symmetric
under spin rotations about the $z$-axis and, in addition, the $z$-component of
the spin is conserved.

The presence of a symmetry leads to conserved quantities and degeneracies.
These, in turn, make it more likely that energy levels will cross as system
parameters are varied. For example, consider a two-level Hamiltonian depending
on a parameter $\lambda\in%
\mathbb{R}
_{+}$ given by
\begin{equation}
\mathrm{H}\left(  \lambda\right)  \overset{\text{def}}{=}\left(
\begin{array}
[c]{cc}%
\lambda & \delta\\
\delta & -\lambda
\end{array}
\right)  \text{.}%
\end{equation}
If $\delta\neq0$, the off-diagonal coupling causes level repulsion (i.e., no
crossing). If $\delta=0$, $\sigma_{z}$ becomes a symmetry of the system and
the energy levels cross at $\lambda=0$. Therefore, one can conclude that level
crossing generally requires decoupling, which is naturally provided by a
symmetry. These factors assist in elucidating the function of symmetry in the
breakdown of quantum search strategies. In particular, one expects that
coupling terms in the search Hamiltonian can help in removing the failures.
Indeed, consider search Hamiltonians of the form \textrm{H}$\left(  t\right)
\overset{\text{def}}{=}\left[  1-s(t)\right]  \mathrm{H}_{0}+s\left(
t\right)  \mathrm{H}_{f}$ with $s\left(  t\right)  \in\left[  0\text{,
}1\right]  $, $\mathrm{H}_{0}\overset{\text{def}}{=}-\left\vert \psi
_{0}\right\rangle \left\langle \psi_{0}\right\vert $, $\mathrm{H}_{f}%
\overset{\text{def}}{=}-\left\vert \psi_{f}\right\rangle \left\langle \psi
_{f}\right\vert $, and $\left\langle \psi_{0}\left\vert \psi_{f}\right.
\right\rangle =0$, the search fails for two reasons. First, the failure occurs
because there is no overlap (i.e., $\left\langle \psi_{f}\left\vert
\mathrm{H}\right\vert \psi_{0}\right\rangle =0$) to allow a direct transition
via the interpolating Hamiltonian $\mathrm{H}$. Second, the failure occurs
because the minimum energy gap vanishes (level crossing) in the presence of
symmetry specified by the commutation relations $\left[  \mathrm{H}\left(
t\right)  \text{, }\mathrm{H}_{0}\right]  =0$ and $\left[  \mathrm{H}\left(
t\right)  \text{, }\mathrm{H}_{f}\right]  =0$ for any $t$. For example,
setting $\left\vert \psi_{0}\right\rangle =\left\vert 0\right\rangle $ and
$\left\vert \psi_{f}\right\rangle =\left\vert 1\right\rangle $, the energy gap
is zero at $s=1/2$. The failure can be fixed by introducing a coupling term
like \textrm{H}$_{\mathrm{coupling}}\overset{\text{def}}{=}\gamma\left(
\left\vert \psi_{0}\right\rangle \left\langle \psi_{f}\right\vert +\left\vert
\psi_{f}\right\rangle \left\langle \psi_{0}\right\vert \right)  $ with
$\gamma>0$ (or, alternatively, designing the search Hamiltonian to act in a
larger subspace). Indeed, considering the Hamiltonian \textrm{H}$\left(
t\right)  \overset{\text{def}}{=}-\left(  1-s\right)  \left\vert
0\right\rangle \left\langle 0\right\vert -s\left\vert 1\right\rangle
\left\langle 1\right\vert -\gamma\left(  \left\vert 0\right\rangle
\left\langle 1\right\vert +\left\vert 1\right\rangle \left\langle 0\right\vert
\right)  $, the energy gap becomes $\Delta\left(  s\right)  \overset
{\text{def}}{=}2\sqrt{(s-1/2)^{2}+\gamma^{2}}$. Therefore, the minimum energy
gap reduces to $\Delta_{\min}=2\gamma>0$ at $s=1/2$. Thus, the search can
succeed in finite time. In general, the analog search Hamiltonian does not
need to be optimal to work, it just needs to ensure non-zero coupling between
the source and target states. If the coupling is weak, the runtime becomes
large. However, the search still succeeds. A tabular summary detailing the
causes of the failure of both the analyzed stationary and nonstationary search
Hamiltonians when the source and target states are orthogonal is presented in
Table II.

In conclusion, drawing from our insights, we determine that in order to
prevent the failure of search algorithms, whether they involve constant or
nonconstant Hamiltonians, it is essential to steer clear of highly symmetric
configurations. Intuitively, this aligns with the understanding that when one
proposes a solution to a problem, it is rare to achieve the ideal solution on
the first try, and variations from the optimal solution are to be expected
during this preliminary phase.

From our conversation regarding failures in quantum searching, a question
arises quite naturally. In the scenario of standard Hamiltonian evolutions
between specified initial and final orthogonal states, what symmetry exists
within the system? What is the symmetry operator that commutes with the
Hamiltonian? It is interesting to point out that in quantum mechanics,
degeneracies arise from inherent symmetry and are removed when that symmetry
is disrupted. For instance, when analyzing a quantum-mechanical particle
confined in a box, the degeneracy exists due to the symmetry that is
established when a general rectangular box is transformed into a cube
\cite{donald}. Subsequently, the degeneracy is eliminated when the symmetry is
compromised by altering the lengths of the sides. When focusing on qubit
dynamics between orthogonal states with a constant Hamiltonian, from a
group-theoretic perspective, it can be stated that there exists a discrete
antipodal symmetry of the projective Hilbert space governed by \textrm{SU}%
$\left(  2;\text{ }%
\mathbb{C}
\right)  $ dynamics defined by stationary Hamiltonians. Specifically, the
antipodal $%
\mathbb{Z}
_{2}$ inversion symmetry of the Bloch sphere is present (or absent) when
considering stationary Hamiltonian evolutions between orthogonal (or
nonorthogonal) states. Here, $%
\mathbb{Z}
_{2}$ denotes the cyclic group of order $2$ formed by two elements, the
identity $e$ and the involution $g$ with $g^{2}=e$. This antipodal symmetry is
expressed by linking the initial and final states through the inversion map
$g:$ $\mathbf{a\rightarrow-a}$, which effectively flips each Bloch vector
$\mathbf{a}$ through the origin. In scenarios involving orthogonal states, the
evolution maintains symmetry concerning the inversion about the sphere's
center. Conversely, when the evolution occurs between nonorthogonal states,
this symmetry is forfeited, resulting in a diminished constraint on the
dynamics. In the orthogonal scenario, since the states are interconnected by
inversion symmetry, this symmetry imposes a constraint: the only constant
Hamiltonian evolution that can precisely map one state to another must adhere
to that symmetry. Generally, $%
\mathbb{Z}
_{2}$ inversions are implemented through unitary involutions, which are
inversion operators $S$ satisfying $S^{2}=\mathbf{1}$, $S\left\vert
A\right\rangle =\left\vert B\right\rangle $, $S\left\vert B\right\rangle
=\left\vert A\right\rangle $, and $\left[  \text{\textrm{H},\textrm{\ }%
}S\right]  =0$. For additional insights regarding group-theoretic arguments in
physics, we suggest that interested readers refer to Ref. \cite{morton}.

With these final comments, we are now prepared to present our summary of
findings and concluding insights. \begin{table}[t]
\centering
\begin{tabular}
[c]{c|c|c|c}\hline\hline
\textbf{Search scheme} & \textbf{Search Hamiltonian} & \textbf{Search failure
and orthogonality} & \textbf{Reason of failure}\\\hline
Farhi-Gutmann & Stationary & Yes & Infinite search time\\\hline
Fenner & Stationary & Yes & Scenario excluded by construction\\\hline
Roland-Cerf & Nonstationary & Yes & Vanishing energy gap\\\hline
\end{tabular}
\caption{Tabular overview outlining the reasons for the failure of both the
examined stationary and nonstationary search Hamiltonians when the source and
target states are orthogonal.}%
\end{table}

\section{Final remarks}

The results outlined in this paper were influenced by two main factors. First,
in the field of analog quantum searching that utilizes time-independent
Hamiltonians, the search technique is ineffective when the source and target
states are orthogonal. Second, with respect to optimal-time evolutions, it is
impossible to evolve in a two-dimensional space in a time suboptimal manner
between two orthogonal states while employing a stationary Hamiltonian. Driven
by these two constraints related to orthogonal states, this paper aimed to
thoroughly examine these matters to enhance our theoretical comprehension of
the shared principles that underpin both the achievements and shortcomings of
analog quantum searches and time-optimal quantum evolutions, based on
fundamental physical principles and symmetry considerations.

\medskip

Our analysis shows that at the level of single-qubit dynamics generated by a
time-independent Hamiltonian, evolution between orthogonal states is strongly
constrained by symmetry. In particular, the dynamics preserves an antipodal
inversion symmetry on the Bloch sphere: orthogonal states lie at opposite
poles, and unitary evolution corresponds to a rotation that necessarily
respects this geometric inversion symmetry. As a result, within this setting,
there is no mechanism to break the symmetry in a way that would qualitatively
alter the structure of the evolution. In the context of quantum search
Hamiltonians, instead, we showed that the presence of continuous symmetries,
such as rotational invariance, frequently gives rise to energy degeneracies
and possible level crossings in the spectrum. Such degeneracies can obstruct
controlled dynamical evolution, for example by preventing the opening of an
energy gap that governs the runtime of the algorithm. By deliberately breaking
certain symmetries- while preserving those essential to the computational
task- one can lift degeneracies, avoid level crossings, and generate a finite
spectral gap. This gap, in turn, ensures well-defined and finite search times.
In this sense, symmetry is not merely a structural constraint but a resource
that can either hinder or enable efficient quantum dynamics, depending on how
it is engineered.

\medskip

An outline of our principal discoveries is as follows:

\begin{enumerate}
\item[{[i]}] We clarified that both \textrm{H}$_{\mathrm{FG}}$ (Eq.
(\ref{FG})) and \textrm{H}$_{\mathrm{Fenner}}$ (Eq. (\ref{Fenner})) are
optimal from the perspective of quantum search ((i.e., in terms of minimum
number of oracle calls)). However, regarding time optimal evolution,
\textrm{H}$_{\mathrm{FG}}$ is deemed suboptimal, while \textrm{H}%
$_{\mathrm{Fenner}}$ is optimal. Additionally, due to its time suboptimal
nature, \textrm{H}$_{\mathrm{FG}}$ is ineffective for searches between
orthogonal states. Conversely, \textrm{H}$_{\mathrm{Fenner}}$ has been
demonstrated to be a time-optimal Hamiltonian (Eq. (\ref{amy})) suitable for
nonorthogonal states. Consequently, it was inherently designed not to be
utilized for searches between orthogonal states (Table II).

\item[{[ii]}] We employed energy arguments (Eq. (\ref{c7})) to demonstrate
that it is not feasible to unitarily evolve between orthogonal states using a
constant Hamiltonian within the two-dimensional subspace defined by the
initial and final states in a manner that is suboptimal in time.

\item[{[iii]}] From a geometric perspective, we illustrated (Fig. $1$ and
Table I) that the evolution between orthogonal states governed by a constant
Hamiltonian is defined by a symmetry determined by the instantaneous
orthogonality conditions between the Bloch vector $\mathbf{a}\left(  t\right)
$ of the evolving state vector $\left\vert \psi\left(  t\right)  \right\rangle
$ and the Bloch vectors of the energy eigenstates $\mathbf{e}_{\pm}$. These
conditions arise from the constancy of the probability amplitudes $\left\vert
\left\langle E_{\pm}\left\vert \psi\left(  t\right)  \right.  \right\rangle
\right\vert ^{2}=1/2$ and are further characterized by the relation
$\mathbf{a}\left(  t\right)  \cdot\mathbf{e}_{\pm}=0$, at any given instant
$t$.

\item[{[iv]}] We investigated a time suboptimal evolution between orthogonal
states utilizing a time-dependent Hamiltonian. We explicitly confirmed that
the violation of the symmetry condition $\mathbf{a}\left(  t\right)
\cdot\mathbf{e}_{\pm}\left(  t\right)  \neq0$ permits a deviation from time
optimality, with the probability amplitude $\left\vert \left\langle E_{\pm
}\left(  t\right)  \left\vert \psi\left(  t\right)  \right.  \right\rangle
\right\vert ^{2}$ typically differing from $1/2$ ( Eqs. (\ref{face4}),
(\ref{face5}), (\ref{face4b}), (\ref{face5b}) and Fig. $2$).

\item[{[v]}] We connected the shortcomings of adiabatic quantum searches (Eq.
(\ref{HRC})) between orthogonal states, attributed to the symmetries of the
time-dependent search Hamiltonian that lead to a vanishing minimum energy gap,
with the failures observed with the transition between orthogonal states with
a constant Hamiltonian in a sub-optimal time fashion.

\medskip
\end{enumerate}

The inability to transition between orthogonal states with a constant
Hamiltonian in a sub-optimal time is significantly related to the failure of
analog quantum search when the source and target states are orthogonal and not
coupled by the Hamiltonian. In both cases, the root of the failure is the
presence of an underlying symmetry of the system. These underlying symmetries
can be geometrically grasped by means of Figs. $1$ and $3$. In Fig. $1$, the
geometry of Bloch vectors for a time optimal evolution (Fig. $1$, $\left(
a\right)  $; time optimal) between orthogonal states displays a higher degree
of symmetry compared with the geometry of Bloch vectors for a time sub-optimal
evolution (Fig. $1$, $(b)$; time sub-optimal) between the same orthogonal
initial and final states. Similarly, Fig. $3$ hints to the fact that the
square with diagonals (Fig. $3$, $\left(  a\right)  $; crossing) exhibits a
higher degree of symmetry than the square with two non-intersecting curves
(Fig. $3$, $\left(  b\right)  $; no crossing).

\medskip

Our analysis in this paper centers on both discrete and continuous symmetries
of quantum systems. In the discrete symmetry setting, we focused on qubit
dynamics between orthogonal states with a constant Hamiltonian. In this case,
to experimentally detect the presence of antipodal $%
\mathbb{Z}
_{2}$ inversion symmetry in a system with Hamiltonian\textbf{ }\textrm{H}%
$\overset{\text{def}}{=}\mathbf{h}\cdot\mathbf{\boldsymbol{\sigma}}$, one can
proceed as follows. Pick the axis $\mathbf{n}$ of the putative $\pi$-flip
symmetry operator $S\overset{\text{def}}{=}\mathbf{n}\cdot
\mathbf{\boldsymbol{\sigma}}$ with $\left[  \mathrm{H}\text{, }e^{-i\frac{\pi
}{2}S}\right]  =2\left(  \mathbf{h\times n}\right)  \cdot
\mathbf{\boldsymbol{\sigma}}$, where $U_{\pi}(\mathbf{n})\overset{\text{def}%
}{=}e^{-i\frac{\pi}{2}S}=e^{-i\frac{\pi}{2}\mathbf{n}\cdot
\mathbf{\boldsymbol{\sigma}}}$ is a $\pi$ rotation about the unit vector
$\mathbf{n}$. Prepare the qubit in an eigenstate of $S$ and let it evolve
under the Hamiltonian $\mathrm{H}$. If the $%
\mathbb{Z}
_{2}$ symmetry is indeed present, the expectation value of $S$ remains
constant over time. In this scenario, $\left\langle S\left(  0\right)
\right\rangle =\left\langle S(t)\right\rangle $ with $\mathbf{h}$ and\textbf{
}$\mathbf{n}$\textbf{ }aligned\textbf{. }Conversely, if the $%
\mathbb{Z}
_{2}$ symmetry is absent, the expectation value of\textbf{ }$S$ will vary over
time. In this case\textbf{, }$\left\langle S\left(  0\right)  \right\rangle
\neq\left\langle S(t)\right\rangle $ with $\mathbf{h}$ and $\mathbf{n}$ not
aligned. In the continuous symmetry setting, we focused on quantum search
Hamiltonians. In this case, we essentially limited our attention to the
invariance of the Hamiltonians under rotations by any angle $\theta$ around an
arbitrary axis $\mathbf{n}$. Comparing discrete and continuous symmetry
settings\textbf{,} $%
\mathbb{Z}
_{2}$ is replaced by $\mathrm{U}(1$; $%
\mathbb{C}
)\overset{\text{def}}{=}\left\{  e^{i\theta}:\theta\in\left[  0\text{, }%
2\pi\right)  \right\}  $ and, in particular, $e^{-i\frac{\theta}{2}%
\mathbf{n}\cdot\mathbf{\boldsymbol{\sigma}}}$\textbf{ }replaces $e^{-i\frac
{\pi}{2}\mathbf{n}\cdot\mathbf{\boldsymbol{\sigma}}}$. For instance, the
commutation relation $\left[  \mathrm{H}\text{, }\sigma_{z}\right]  =0$
indicates the presence of an exact, continuous, unitary, global internal
symmetry, corresponding to invariance under spin rotations about the $z$-axis.
From an experimental perspective, we stress that such types of continuous
symmetries are recognized indirectly via conserved observables, selection
rules in spectroscopy, symmetry-protected degeneracies, and dynamics occurring
within invariant subspaces \cite{bertolucci89}. A particularly effective
diagnostic approach is the intentional breaking of symmetry: previously
conserved quantities begin to vary, transitions that were once prohibited
arise, and degeneracies split into avoided crossings. The robustness of these
indicators in the face of parameter variations provides concrete evidence of
the underlying symmetry. For example, to verify that \textrm{H} commutes with
an operator $\mathcal{O}$, one can prepare quantum states with different
initial values of\textbf{ }$\left\langle \mathcal{O(}0\mathcal{)}\right\rangle
$,\textbf{ }let the system evolve, and measure\textbf{ }$\left\langle
\mathcal{O(}t\mathcal{)}\right\rangle $ at later times. If the symmetry is
present, $\left\langle \mathcal{O(}t\mathcal{)}\right\rangle $ remains
constant and equal to $\left\langle \mathcal{O(}0\mathcal{)}\right\rangle $.
Alternatively, using controlled symmetry breaking procedures, one can
introduce a small, tunable perturbation that explicitly breaks the suspected
symmetry (for example, adding a transverse field if $\left[  \mathrm{H}\text{,
}\sigma_{z}\right]  =0$). By observing how the spectrum and dynamics respond,
one can determine the presence of the original symmetry. If conserved
quantities begin to evolve, or if degeneracies split into avoided crossings or
previously forbidden transitions become allowed, the underlying symmetry can
be identified. For more details on theoretical and experimental aspects of
discrete and continuous symmetries beyond Schr\"{o}dinger's evolutions and
quantum search Hamiltonians, we recommend consulting Ref. \cite{bertolucci89}.

\medskip

Summing up, symmetry can be utilized to elucidate the shortcomings of analog
quantum search schemes, whether stationary or nonstationary, in the context of
orthogonal (known) source and (unknown) target states. Furthermore, symmetry
arguments can be employed to demonstrate the impossibility of creating
suboptimal time stationary Hamiltonian evolutions between specified initial
and final orthogonal quantum states. In the first instance, symmetry results
in energy level crossing, which consequently leads to asymptotically prolonged
search times, rendering the search ineffective. Conversely, in the latter
scenario, the existence of a discrete antipodal symmetry prevents the
evolution from deviating from the optimal time trajectory defined by a
geodesic path on the Bloch sphere that connects the initial and final states.

\medskip

We remark that the growing fascination with the significance of symmetries and
conserved quantities in quantum control theory \cite{albertini18} and quantum
search \cite{wang21,meyer25} renders our research particularly appealing to
scientists who are tackling these challenges (i.e., control and search) from a
group-theoretic perspective. Our analysis centered on eliminating failures (in
both search and control) within two-dimensional subspaces of the complete
Hilbert space. However, in the realm of time-optimal evolutions, it is
possible to explore the construction of stationary Hamiltonians that dictate
the evolution between two orthogonal states in higher-dimensional subspaces in
a less than optimal time manner. Furthermore, regarding quantum search between
orthogonal source and target states, one might consider the potential for
failure reduction by increasing the dimensionality of the search space
\cite{popescu24,wang25}, albeit at the cost of optimal query complexity. We
will defer these discussions to future scientific endeavors.

\begin{acknowledgments}
C.C. is grateful to the Griffiss Institute (Rome-NY) and to the United States
Air Force Research Laboratory (AFRL) Visiting Faculty Research Program (VFRP)
for providing support for this work. J.S. gratefully acknowledges support from
the AFRL. The authors express their gratitude for the enlightening discussions
held with P. M. Alsing. Any opinions, findings and conclusions or
recommendations expressed in this material are those of the authors and do not
necessarily reflect the views of the AFRL. The authors thank the three
anonymous referees for very useful comments leading to an improved version of
this manuscript\textbf{.}
\end{acknowledgments}

\bigskip\pagebreak

\appendix

\section{Derivation of Eq. (\ref{PFenner})}

In this appendix, we calculate the transition probability $\mathcal{P}%
_{\mathrm{Fenner}}\left(  t\right)  $ in Eq. (\ref{PFenner}).

Consider the Hamiltonian \textrm{H}$_{\mathrm{Fenner}}\overset{\text{def}}%
{=}2iEx\left(  \left\vert w\right\rangle \left\langle s\right\vert -\left\vert
s\right\rangle \left\langle w\right\vert \right)  $ with the source state
$\left\vert s\right\rangle =x\left\vert w\right\rangle +\sqrt{1-x^{2}%
\left\vert r\right\rangle }$ in the orthonormal basis $\left\{  \left\vert
w\right\rangle \text{, }\left\vert r\right\rangle \right\}  $. Then, the
matrix representation of \textrm{H}$_{\mathrm{Fenner}}$ in the basis $\left\{
\left\vert w\right\rangle \text{, }\left\vert r\right\rangle \right\}  $ is
given by%
\begin{equation}
\left[  \mathrm{H}_{\mathrm{Fenner}}\right]  _{\left\{  \left\vert
w\right\rangle \text{, }\left\vert r\right\rangle \right\}  }\overset
{\text{def}}{=}\left(
\begin{array}
[c]{cc}%
\left\langle w\left\vert \mathrm{H}_{\mathrm{Fenner}}\right\vert
w\right\rangle  & \left\langle w\left\vert \mathrm{H}_{\mathrm{Fenner}%
}\right\vert r\right\rangle \\
\left\langle r\left\vert \mathrm{H}_{\mathrm{Fenner}}\right\vert
w\right\rangle  & \left\langle r\left\vert \mathrm{H}_{\mathrm{Fenner}%
}\right\vert r\right\rangle
\end{array}
\right)  =\left(
\begin{array}
[c]{cc}%
0 & 2iEx\sqrt{1-x^{2}}\\
-2iEx\sqrt{1-x^{2}} & 0
\end{array}
\right)  \text{.}%
\end{equation}
From a simple calculation, we get that the energy eigenvalues of $\left[
\mathrm{H}_{\mathrm{Fenner}}\right]  _{\left\{  \left\vert w\right\rangle
\text{, }\left\vert r\right\rangle \right\}  }$ are $E_{1}\overset{\text{def}%
}{=}2Ex\sqrt{1-x^{2}}$ and $E_{2}\overset{\text{def}}{=}-2Ex\sqrt{1-x^{2}}$. A
corresponding set of orthonormal eigenstates is given by%
\begin{equation}
\left\vert E_{1}\right\rangle \overset{\text{def}}{=}\frac{\left\vert
w\right\rangle -i\left\vert r\right\rangle }{\sqrt{2}}\text{, and }\left\vert
E_{2}\right\rangle \overset{\text{def}}{=}\frac{\left\vert w\right\rangle
+i\left\vert r\right\rangle }{\sqrt{2}}\text{,} \label{eigen2}%
\end{equation}
respectively. Using Eq. (\ref{eigen2}), we can recast the source state, the
target state, and the Hamiltonian \textrm{H}$_{\mathrm{Fenner}}$ in terms of
the eigenstates $\left\vert E_{1}\right\rangle $ and $\left\vert
E_{2}\right\rangle $. We have,%
\begin{equation}
\left\vert s\right\rangle =\frac{x+i\sqrt{1-x^{2}}}{\sqrt{2}}\left\vert
E_{1}\right\rangle +\frac{x-i\sqrt{1-x^{2}}}{\sqrt{2}}\left\vert
E_{2}\right\rangle \text{, }\left\vert w\right\rangle =\frac{\left\vert
E_{1}\right\rangle +\left\vert E_{2}\right\rangle }{\sqrt{2}}\text{,}
\label{loveme1}%
\end{equation}
and, finally,
\begin{equation}
\mathrm{H}_{\mathrm{Fenner}}=E_{1}\left\vert E_{1}\right\rangle \left\langle
E_{1}\right\vert +E_{2}\left\vert E_{2}\right\rangle \left\langle
E_{2}\right\vert \text{.} \label{loveme2}%
\end{equation}
Finally, making use of Eqs. (\ref{loveme1}) and (\ref{loveme2}), the
transition probability $\mathcal{P}_{\mathrm{Fenner}}\left(  t\right)
\overset{\text{def}}{=}\left\vert \left\langle w\left\vert e^{-\frac
{i}{\hslash}\mathrm{H}_{\mathrm{Fenner}}t}\right\vert s\right\rangle
\right\vert ^{2}$ becomes%
\begin{equation}
\mathcal{P}_{\mathrm{Fenner}}\left(  t\right)  =\left\vert x\cos\left(
2x\sqrt{1-x^{2}}\frac{E}{\hslash}t\right)  +\sqrt{1-x^{2}}\sin\left(
2x\sqrt{1-x^{2}}\frac{E}{\hslash}t\right)  \right\vert ^{2}\text{.}
\label{PTT}%
\end{equation}
The calculation of $\mathcal{P}_{\mathrm{Fenner}}\left(  t\right)  $ in Eq.
(\ref{PTT}) ends our derivation.

\section{Quantum search as quantum simulation}

In this appendix, we explain how the time evolution sub-optimality of the
Fahri-Gutmann search Hamiltonian \textrm{H}$_{\mathrm{FG}}=E\left\vert
w\right\rangle \left\langle w\right\vert +E\left\vert s\right\rangle
\left\langle s\right\vert $ can be understood in geometric terms by
re-deriving the search Hamiltonian from the point of view of quantum
simulation. This statement was reported in the end of Section II.

Specifically, one needs to study in an explicit manner the action of a
simulation step of time $\Delta t$ specified by the unitary operator $U(t)$
given by%
\begin{equation}
U(\Delta t)\overset{\text{def}}{=}e^{-i\left\vert s\right\rangle \left\langle
s\right\vert \Delta t}e^{-i\left\vert w\right\rangle \left\langle w\right\vert
\Delta t}\text{,} \label{need1}%
\end{equation}
where we set $E=1=\hslash$. To calculate $U(\Delta t)$ in Eq. (\ref{need1}),
we can focus on studying how one can express the sequential application of two
rotations $\mathcal{R}_{\mathbf{n}_{1}}\left(  \theta_{1}\right)  $ and
$\mathcal{R}_{\mathbf{n}_{2}}\left(  \theta_{2}\right)  $ on a single qubit in
terms of a single rotation specified by $\mathcal{R}_{\mathbf{n}_{1}}\left(
\theta_{1}\right)  \mathcal{R}_{\mathbf{n}_{2}}\left(  \theta_{2}\right)
=\mathcal{R}_{\mathbf{n}}\left(  \theta\right)  $. Formally, setting%
\begin{equation}
e^{-i\frac{\theta_{1}}{2}\mathbf{n}_{1}\cdot\mathbf{\boldsymbol{\sigma}}%
}e^{-i\frac{\theta_{2}}{2}\mathbf{n}_{2}\cdot\mathbf{\boldsymbol{\sigma}}%
}=e^{-i\frac{\theta}{2}\mathbf{n}\cdot\mathbf{\boldsymbol{\sigma}}}\text{,}
\label{be1}%
\end{equation}
one needs to find the angle $\theta$ and the axis of rotation $\mathbf{n}$ in
Eq. (\ref{be1}). We proceed as follows. Note that \cite{nielsen10},%
\begin{align}
e^{-i\frac{\theta}{2}\mathbf{n}\cdot\mathbf{\boldsymbol{\sigma}}}  &
=e^{-i\frac{\theta_{1}}{2}\mathbf{n}_{1}\cdot\mathbf{\boldsymbol{\sigma}}%
}e^{-i\frac{\theta_{2}}{2}\mathbf{n}_{2}\cdot\mathbf{\boldsymbol{\sigma}}%
}\nonumber\\
&  =\left[  \cos\left(  \frac{\theta_{1}}{2}\right)  \mathrm{I}-i\sin\left(
\frac{\theta_{1}}{2}\right)  \mathbf{n}_{1}\cdot\mathbf{\boldsymbol{\sigma}%
}\right]  \left[  \cos\left(  \frac{\theta_{2}}{2}\right)  \mathrm{I}%
-i\sin\left(  \frac{\theta_{2}}{2}\right)  \mathbf{n}_{2}\cdot
\mathbf{\boldsymbol{\sigma}}\right] \nonumber\\
&  =\cos\left(  \frac{\theta_{1}}{2}\right)  \cos\left(  \frac{\theta_{2}}%
{2}\right)  \mathrm{I}-i\cos\left(  \frac{\theta_{1}}{2}\right)  \sin\left(
\frac{\theta_{2}}{2}\right)  \mathbf{n}_{2}\cdot\mathbf{\boldsymbol{\sigma}%
}-i\sin\left(  \frac{\theta_{1}}{2}\right)  \cos\left(  \frac{\theta_{2}}%
{2}\right)  \mathbf{n}_{1}\cdot\mathbf{\boldsymbol{\sigma}}+\nonumber\\
&  -\sin\left(  \frac{\theta_{1}}{2}\right)  \sin\left(  \frac{\theta_{2}}%
{2}\right)  \left(  \mathbf{n}_{1}\cdot\mathbf{\boldsymbol{\sigma}}\right)
\left(  \mathbf{n}_{2}\cdot\mathbf{\boldsymbol{\sigma}}\right)  \text{.}
\label{primo}%
\end{align}
To simplify Eq. (\ref{primo}), we use ordinary quantum-mechanical properties
of Pauli matrices. We have,
\begin{align}
\left(  \mathbf{n}_{1}\cdot\mathbf{\boldsymbol{\sigma}}\right)  \left(
\mathbf{n}_{2}\cdot\mathbf{\boldsymbol{\sigma}}\right)   &  =\left(
n_{1i}\sigma_{i}\right)  \left(  n_{2j}\sigma_{j}\right) \nonumber\\
&  =n_{1i}n_{2j}\sigma_{i}\sigma_{j}\nonumber\\
&  =n_{1i}n_{2j}\left(  \frac{1}{2}\left[  \sigma_{i}\text{, }\sigma
_{j}\right]  +\frac{1}{2}\left\{  \sigma_{i}\text{, }\sigma_{j}\right\}
\right) \nonumber\\
&  =n_{1i}n_{2j}\left(  \frac{1}{2}2i\varepsilon_{ijk}\sigma_{k}+\frac{1}%
{2}2\delta_{ij}\right) \nonumber\\
&  =n_{1i}n_{2j}\delta_{ij}+i\varepsilon_{ijk}n_{1i}n_{2j}\sigma
_{k}\nonumber\\
&  =\left(  \mathbf{n}_{1}\cdot\mathbf{n}_{2}\right)  \mathrm{I}+i\left(
\mathbf{n}_{1}\times\mathbf{n}_{2}\right)  \cdot\mathbf{\boldsymbol{\sigma}%
}\text{,}%
\end{align}
that is,
\begin{equation}
\left(  \mathbf{n}_{1}\cdot\mathbf{\boldsymbol{\sigma}}\right)  \left(
\mathbf{n}_{2}\cdot\mathbf{\boldsymbol{\sigma}}\right)  =\left(
\mathbf{n}_{1}\cdot\mathbf{n}_{2}\right)  \mathrm{I}+i\left(  \mathbf{n}%
_{1}\times\mathbf{n}_{2}\right)  \cdot\mathbf{\boldsymbol{\sigma}}\text{.}
\label{secondo}%
\end{equation}
Therefore, substituting Eq. (\ref{secondo}) into Eq. (\ref{primo}), we arrive
at%
\begin{align}
e^{-i\frac{\theta}{2}\mathbf{n}\cdot\mathbf{\boldsymbol{\sigma}}}  &
=\cos\left(  \frac{\theta_{1}}{2}\right)  \cos\left(  \frac{\theta_{2}}%
{2}\right)  \mathrm{I}-i\cos\left(  \frac{\theta_{1}}{2}\right)  \sin\left(
\frac{\theta_{2}}{2}\right)  \mathbf{n}_{2}\cdot\mathbf{\boldsymbol{\sigma}%
}-i\sin\left(  \frac{\theta_{1}}{2}\right)  \cos\left(  \frac{\theta_{2}}%
{2}\right)  \mathbf{n}_{1}\cdot\mathbf{\boldsymbol{\sigma}}+\nonumber\\
&  -\sin\left(  \frac{\theta_{1}}{2}\right)  \sin\left(  \frac{\theta_{2}}%
{2}\right)  \left(  \mathbf{n}_{1}\cdot\mathbf{n}_{2}\right)  \mathrm{I}%
-i\sin\left(  \frac{\theta_{1}}{2}\right)  \sin\left(  \frac{\theta_{2}}%
{2}\right)  \left(  \mathbf{n}_{1}\times\mathbf{n}_{2}\right)  \cdot
\mathbf{\boldsymbol{\sigma}}\nonumber\\
& \nonumber\\
&  =\left[  \cos\left(  \frac{\theta_{1}}{2}\right)  \cos\left(  \frac
{\theta_{2}}{2}\right)  -\sin\left(  \frac{\theta_{1}}{2}\right)  \sin\left(
\frac{\theta_{2}}{2}\right)  \left(  \mathbf{n}_{1}\cdot\mathbf{n}_{2}\right)
\right]  \mathrm{I}+\nonumber\\
&  -i\left[  \sin\left(  \frac{\theta_{1}}{2}\right)  \cos\left(  \frac
{\theta_{2}}{2}\right)  \mathbf{n}_{1}+\cos\left(  \frac{\theta_{1}}%
{2}\right)  \sin\left(  \frac{\theta_{2}}{2}\right)  \mathbf{n}_{2}%
-\sin\left(  \frac{\theta_{1}}{2}\right)  \sin\left(  \frac{\theta_{2}}%
{2}\right)  \left(  \mathbf{n}_{2}\times\mathbf{n}_{1}\right)  \right]
\cdot\mathbf{\boldsymbol{\sigma}}\nonumber\\
& \nonumber\\
&  =\left[  \cos\left(  \frac{\theta}{2}\right)  \mathrm{I}-i\sin\left(
\frac{\theta}{2}\right)  \mathbf{n}\cdot\mathbf{\boldsymbol{\sigma}}\right]
\text{,}%
\end{align}
that is,%
\begin{equation}
\left\{
\begin{array}
[c]{c}%
\cos\left(  \frac{\theta}{2}\right)  =\cos\left(  \frac{\theta_{1}}{2}\right)
\cos\left(  \frac{\theta_{2}}{2}\right)  -\sin\left(  \frac{\theta_{1}}%
{2}\right)  \sin\left(  \frac{\theta_{2}}{2}\right)  \left(  \mathbf{n}%
_{1}\cdot\mathbf{n}_{2}\right)  \text{,}\\
\\
\sin\left(  \frac{\theta}{2}\right)  \mathbf{n}=\sin\left(  \frac{\theta_{1}%
}{2}\right)  \cos\left(  \frac{\theta_{2}}{2}\right)  \mathbf{n}_{1}%
+\cos\left(  \frac{\theta_{1}}{2}\right)  \sin\left(  \frac{\theta_{2}}%
{2}\right)  \mathbf{n}_{2}-\sin\left(  \frac{\theta_{1}}{2}\right)
\sin\left(  \frac{\theta_{2}}{2}\right)  \left(  \mathbf{n}_{2}\times
\mathbf{n}_{1}\right)
\end{array}
\right.  \text{.} \label{about}%
\end{equation}
Finally, Eq. (\ref{about}) allows us to specify the rotation angle $\theta$
and the rotation axis $\mathbf{n}$ in $\mathcal{R}_{\mathbf{n}}\left(
\theta\right)  =\mathcal{R}_{\mathbf{n}_{1}}\left(  \theta_{1}\right)
\mathcal{R}_{\mathbf{n}_{2}}\left(  \theta_{2}\right)  $.

Returning to the expression for $U(\Delta t)$ in Eq. (\ref{need1}), let us
set
\begin{equation}
\left\vert s\right\rangle \left\langle s\right\vert =\frac{\mathrm{I}%
+\mathbf{s}\cdot\mathbf{\boldsymbol{\sigma}}}{2}\text{, and }\left\vert
w\right\rangle \left\langle w\right\vert =\frac{\mathrm{I}+\mathbf{w}%
\cdot\mathbf{\boldsymbol{\sigma}}}{2}\text{,} \label{jesus}%
\end{equation}
where $\mathbf{s}\overset{\text{def}}{\mathbf{=}}(2x\sqrt{1-x^{2}}$, $0$,
$2x^{2}-1)$ and $\mathbf{w}\overset{\text{def}}{\mathbf{=}}\hat{z}=(0$, $0$,
$1)$ are unit vectors. Substituting Eq. (\ref{jesus}) into Eq. (\ref{need1}),
$U\left(  \Delta t\right)  $ becomes%
\begin{align}
U\left(  \Delta t\right)   &  =e^{-i\left\vert s\right\rangle \left\langle
s\right\vert \Delta t}e^{-i\left\vert w\right\rangle \left\langle w\right\vert
\Delta t}\nonumber\\
&  =e^{-i\frac{\mathrm{I}+\mathbf{s}\cdot\mathbf{\boldsymbol{\sigma}}}%
{2}\Delta t}e^{-i\frac{\mathrm{I}+\mathbf{w}\cdot\mathbf{\boldsymbol{\sigma}}%
}{2}\Delta t}\nonumber\\
&  =e^{-i\frac{\mathrm{I}}{2}\Delta t}e^{-i\frac{\mathbf{s}\cdot
\mathbf{\boldsymbol{\sigma}}}{2}\Delta t}e^{-i\frac{\mathrm{I}}{2}\Delta
t}e^{-i\frac{\mathbf{w}\cdot\mathbf{\boldsymbol{\sigma}}}{2}\Delta
t}\nonumber\\
&  =e^{-i\mathrm{I}\Delta t}e^{-i\frac{\mathbf{s}\cdot
\mathbf{\boldsymbol{\sigma}}}{2}\Delta t}e^{-i\frac{\mathbf{w}\cdot
\mathbf{\boldsymbol{\sigma}}}{2}\Delta t}\nonumber\\
&  \approx e^{-i\frac{\mathbf{s}\cdot\mathbf{\boldsymbol{\sigma}}}{2}\Delta
t}e^{-i\frac{\mathbf{w}\cdot\mathbf{\boldsymbol{\sigma}}}{2}\Delta t}\text{,}
\label{tomen}%
\end{align}
where in the last step in Eq. (\ref{tomen}) we neglect the irrelevant global
phase factor $e^{-i\mathrm{I}\Delta t}$. Now, we want to use Eq. (\ref{about})
to express $e^{-i\frac{\mathbf{s}\cdot\mathbf{\boldsymbol{\sigma}}}{2}\Delta
t}e^{-i\frac{\mathbf{w}\cdot\mathbf{\boldsymbol{\sigma}}}{2}\Delta t}$ as
$e^{-i\frac{\theta}{2}\mathbf{n}\cdot\mathbf{\boldsymbol{\sigma}}}$. Using
Eqs. (\ref{about}) and (\ref{tomen}), we find that $\mathbf{n}$ and $\theta$
such that $e^{-i\frac{\theta}{2}\mathbf{n}\cdot\mathbf{\boldsymbol{\sigma}}%
}=e^{-i\frac{\mathbf{s}\cdot\mathbf{\boldsymbol{\sigma}}}{2}\Delta
t}e^{-i\frac{\mathbf{w}\cdot\mathbf{\boldsymbol{\sigma}}}{2}\Delta t}$ satisfy
the following two relations,%
\begin{equation}
\left\{
\begin{array}
[c]{c}%
\cos\left(  \frac{\theta}{2}\right)  =\cos^{2}\left(  \frac{\Delta t}%
{2}\right)  -\sin^{2}\left(  \frac{\Delta t}{2}\right)  \mathbf{s}%
\cdot\mathbf{w}\\
\sin\left(  \frac{\theta}{2}\right)  \mathbf{n}=\sin\left(  \frac{\Delta t}%
{2}\right)  \cos\left(  \frac{\Delta t}{2}\right)  \left(  \mathbf{s}%
+\mathbf{w}\right)  -\sin^{2}\left(  \frac{\Delta t}{2}\right)  \mathbf{w}%
\times\mathbf{s}%
\end{array}
\right.  \text{.} \label{tomen2}%
\end{equation}
Therefore, from Eqs. (\ref{tomen}) and (\ref{tomen2}), $U\left(  \Delta
t\right)  \approx e^{-i\frac{\mathbf{s}\cdot\mathbf{\boldsymbol{\sigma}}}%
{2}\Delta t}e^{-i\frac{\mathbf{w}\cdot\mathbf{\boldsymbol{\sigma}}}{2}\Delta
t}$ can be recast as%
\begin{equation}
U\left(  \Delta t\right)  \approx\left[  \cos^{2}\left(  \frac{\Delta t}%
{2}\right)  -\sin^{2}\left(  \frac{\Delta t}{2}\right)  \mathbf{s}%
\cdot\mathbf{w}\right]  \mathrm{I}-2i\sin\left(  \frac{\Delta t}{2}\right)
\left[  \cos\left(  \frac{\Delta t}{2}\right)  \left(  \frac{\mathbf{s}%
+\mathbf{w}}{2}\right)  +\sin\left(  \frac{\Delta t}{2}\right)  \left(
\frac{\mathbf{s}\times\mathbf{w}}{2}\right)  \right]  \cdot
\mathbf{\boldsymbol{\sigma}}\text{.} \label{omer}%
\end{equation}
The unnormalized axis of rotation that specifies $U\left(  \Delta t\right)  $
in Eq. (\ref{omer}) is given by%
\begin{equation}
\mathbf{r}\overset{\text{def}}{=}\cos\left(  \frac{\Delta t}{2}\right)
\left(  \frac{\mathbf{s}+\mathbf{w}}{2}\right)  +\sin\left(  \frac{\Delta
t}{2}\right)  \left(  \frac{\mathbf{s}\times\mathbf{w}}{2}\right)  \text{,}
\label{rdef}%
\end{equation}
with $\mathbf{r}=\left\Vert \mathbf{r}\right\Vert \hat{r}=(x\sqrt{1-x^{2}%
})\cos(\Delta t/2)$, $-x\sqrt{1-x^{2}})\sin(\Delta t/2)$, $x^{2}\cos\left(
\Delta t/2\right)  $. After some algebra with Eqs. (\ref{tomen2}),
(\ref{omer}), and (\ref{rdef}), $U\left(  \Delta t\right)  $ can be finally
expressed as%
\begin{equation}
U\left(  \Delta t\right)  =\cos\left[  \frac{\theta\left(  t\right)  }%
{2}\right]  \mathrm{I}-i\sin\left[  \frac{\theta\left(  t\right)  }{2}\right]
\hat{r}\left(  \Delta t\right)  \cdot\mathbf{\boldsymbol{\sigma}}\text{,}
\label{UT}%
\end{equation}
where $\hat{r}\left(  \Delta t\right)  \overset{\text{def}}{=}\mathbf{r}%
\left(  \Delta t\right)  /\left\Vert \mathbf{r}\left(  \Delta t\right)
\right\Vert $ with $\mathbf{r=r}\left(  \Delta t\right)  $ in Eq. (\ref{rdef})
and, in addition,
\begin{equation}
\cos\left[  \frac{\theta\left(  t\right)  }{2}\right]  =1-\frac{2}{N}\sin
^{2}\left(  \frac{\Delta t}{2}\right)  \text{, and }\sin\left[  \frac
{\theta\left(  t\right)  }{2}\right]  =\frac{2}{\sqrt{N}}\sin(\frac{\Delta
t}{2})\sqrt{1-\frac{1}{N}\sin^{2}(\frac{\Delta t}{2})}\text{,}%
\end{equation}
with $x=\left\vert \left\langle s\left\vert w\right.  \right\rangle
\right\vert =1/\sqrt{N}$ and $N$ being the dimensionality of the search space.
It is interesting to point out that since $\mathbf{r}\cdot\mathbf{s}%
=\mathbf{r}\cdot\mathbf{w}=x^{2}\cos\left(  \Delta t/2\right)  $, both
$\left\vert s\right\rangle \left\langle s\right\vert $ and $\left\vert
w\right\rangle \left\langle w\right\vert $ lie on the same circle of
revolution about the $\mathbf{r}$-axis on the Bloch sphere. In summary, the
action of each application of $U\left(  \Delta t\right)  $ is to rotate
$\left\vert s\right\rangle \left\langle s\right\vert $ by an angle $\theta$
about the $\mathbf{r}$-axis. The procedure ends when enough rotations have
been performed so that $\left\vert s\right\rangle \left\langle s\right\vert $
is near the solution $\left\vert w\right\rangle \left\langle w\right\vert $.
Of course, there are multiple distinct ways of arriving at $\left\vert
w\right\rangle \left\langle w\right\vert $ from $\left\vert s\right\rangle
\left\langle s\right\vert $, depending on the choice of the simulation step
length $\Delta t$ that specifies the rotation angle $\theta\left(  \Delta
t\right)  $ and the rotation axis $\hat{r}\left(  \Delta t\right)  $ given by%
\begin{equation}
\hat{r}\left(  \Delta t\right)  =\frac{\cos\left(  \frac{\Delta t}{2}\right)
\left(  \frac{\mathbf{s}+\mathbf{w}}{2}\right)  +\sin\left(  \frac{\Delta
t}{2}\right)  \left(  \frac{\mathbf{s}\times\mathbf{w}}{2}\right)
}{\left\Vert \cos\left(  \frac{\Delta t}{2}\right)  \left(  \frac
{\mathbf{s}+\mathbf{w}}{2}\right)  +\sin\left(  \frac{\Delta t}{2}\right)
\left(  \frac{\mathbf{s}\times\mathbf{w}}{2}\right)  \right\Vert }\text{.}
\label{vedivedi}%
\end{equation}
From Eq. (\ref{vedivedi}) we note that $\hat{r}\left(  \Delta t\right)  $
belongs to the space spanned by $\left\{  \mathbf{s}+\mathbf{w}\text{,
}\mathbf{s}\times\mathbf{w}\right\}  $. However, as reported in Refs.
\cite{rossetti24,rossetti25}, the optimal rotation would be the one in which
$\hat{r}\left(  \Delta t\right)  $ belongs to the space spanned by $\left\{
\mathbf{s}\times\mathbf{w}\right\}  $. This, in turn, would imply choosing a
$\Delta t=\pi$ in Eq. (\ref{vedivedi}). With this choice, the quantum
simulation specified by $U\left(  \Delta t\right)  $ in Eq. (\ref{need1})
equals $U\left(  \pi\right)  =e^{-i\pi\left\vert s\right\rangle \left\langle
s\right\vert }e^{-i\pi\left\vert w\right\rangle \left\langle w\right\vert
\Delta t}$ with $e^{-i\pi\left\vert s\right\rangle \left\langle s\right\vert
}=-\mathbf{s\cdot\boldsymbol{\sigma=}}\mathrm{I}-2\left\vert s\right\rangle
\left\langle s\right\vert $ and $e^{-i\pi\left\vert w\right\rangle
\left\langle w\right\vert }=-\mathbf{w\cdot\boldsymbol{\sigma=}}%
\mathrm{I}-2\left\vert w\right\rangle \left\langle w\right\vert $. Therefore,
modulo a global phase shift, $U\left(  \pi\right)  $ is identical to Grover's
iterate in a digital quantum search setting. With this final remark, we end
our discussion here.

\section{Derivation of Eq. (\ref{lunghezza})}

In this appendix, we present fundamental elements of the Fubini-Study metric
to facilitate the understanding of how to compute path lengths (i.e., Eq.
(\ref{lunghezza})) in projective Hilbert space.

Let $\mathcal{H}=\mathcal{H}_{2}^{n}$ be an $N\overset{\text{def}}{=}2^{n}%
$-dimensional complex Hilbert space of $n$-qubit unit quantum states $\left\{
\left\vert \psi\right\rangle \right\}  $. Given the fact that the global phase
of a vector state cannot be observed, a physical state can be represented by a
so-called ray of the Hilbert space. The set of rays of $\mathcal{H}_{2}^{n}$
\ defines the so-called complex projective Hilbert space $\mathcal{P(H)=}%
\mathbb{C}
P^{N-1}$. More formally, $%
\mathbb{C}
P^{N-1}$ denotes the quotient set of $\mathcal{H}_{2}^{n}$ by the equivalence
relation $\left\vert \psi\right\rangle \sim e^{i\phi}\left\vert \psi
\right\rangle $ with $\phi\in%
\mathbb{R}
$. The space $%
\mathbb{C}
P^{N-1}$ can be endowed with a mathematically valid and physically significant
metric structure. In fact, let us consider a family $\left\{  \left\vert
\psi\left(  \xi\right)  \right\rangle \right\}  $ of normalized quantum states
of $\mathcal{H}_{2}^{n}$ that vary smoothly with respect to an $m$-dimensional
parameter $\xi\overset{\text{def}}{=}\left(  \xi^{1}\text{,..., }\xi
^{m}\right)  \in%
\mathbb{R}
^{m}$. Consequently, the standard Hermitian scalar product on $\mathcal{H}%
_{2}^{n}$ gives rise to a metric tensor $g_{ab}\left(  \xi\right)  $ with
$1\leq a$, $b\leq m$ on the manifold of quantum states as defined as
\cite{provost80},%
\begin{equation}
g_{ab}\left(  \xi\right)  \overset{\text{def}}{=}4\operatorname{Re}\left[
\left\langle \partial_{a}\psi\left(  \xi\right)  |\partial_{b}\psi\left(
\xi\right)  \right\rangle -\left\langle \partial_{a}\psi\left(  \xi\right)
|\psi\left(  \xi\right)  \right\rangle \left\langle \psi\left(  \xi\right)
|\partial_{b}\psi\left(  \xi\right)  \right\rangle \right]  \text{,}
\label{fubini}%
\end{equation}
where $\partial_{a}\overset{\text{def}}{=}\partial/\partial\xi^{a}$. The
quantity $g_{ab}\left(  \xi\right)  $ in Eq. (\ref{fubini}) represents the
so-called Fubini-Study metric tensor. Notably, we observe that this metric is
positive definite, which is clear when examining the distance element
$ds_{\text{\textrm{FS}}}^{2}$ between two adjacent points corresponding to the
vector states $\left\vert \psi\left(  \xi+d\xi\right)  \right\rangle $ and
$\left\vert \psi\left(  \xi\right)  \right\rangle $ \cite{brau94},%
\begin{equation}
ds_{\text{\textrm{FS}}}^{2}\overset{\text{def}}{=}g_{ab}\left(  \xi\right)
d\xi^{a}d\xi^{b}=4\left[  \left\langle d\psi|d\psi\right\rangle -\left\vert
\left\langle \psi|d\psi\right\rangle \right\vert ^{2}\right]  \text{,}
\label{distance1}%
\end{equation}
with $\left\vert d\psi\right\rangle \overset{\text{def}}{=}\left\vert
\psi\left(  \xi+d\xi\right)  \right\rangle -\left\vert \psi\left(  \xi\right)
\right\rangle $. The distance element $ds_{\text{\textrm{FS}}}^{2}$ in Eq.
(\ref{distance1}) naturally introduces the idea of geodesic paths in $%
\mathbb{C}
P^{N-1}$. For more details on the notion of Fubini-Study metric, we suggest
Ref. \cite{provost80}.

To calculate path lengths, one needs to use $g_{ab}\left(  \xi\right)  $ in
Eq. (\ref{fubini}). In particular, it can be shown that the quantum line
between states $\left\vert A\right\rangle $ and $\left\vert B\right\rangle $
can be parametrized in terms of a single parameter $\xi\in\left[  0\text{,
}\pi\right]  $ as \cite{cafarocqg},%
\begin{equation}
\left\vert \psi\left(  \xi\right)  \right\rangle =\frac{\cos\left(  \frac{\xi
}{2}\right)  \left\vert A\right\rangle +\frac{\left\langle B\left\vert
A\right.  \right\rangle }{\left\vert \left\langle B\left\vert A\right.
\right\rangle \right\vert }\sin(\frac{\xi}{2})\left\vert B\right\rangle
}{\sqrt{1+\sin\left(  \xi\right)  \left\vert \left\langle B\left\vert
A\right.  \right\rangle \right\vert }}\text{,} \label{carluccio}%
\end{equation}
with $\left\vert \psi\left(  0\right)  \right\rangle =\left\vert
A\right\rangle $ and $\left\vert \psi\left(  \pi\right)  \right\rangle
\simeq\left\vert B\right\rangle $ (with \textquotedblleft$\simeq
$\textquotedblright\ denoting physical equivalence between quantum states,
modulo irrelevant global phase factors). The parameter $\xi$ is related to the
temporal parameter $t$ (adjusted in such a manner that $0\leq t\leq1$) by
means of the relation $t\left(  \xi\right)  \overset{\text{def}}{=}\tan\left(
\xi/2\right)  /\left[  1+\tan(\xi/2)\right]  $ so that $t\left(  0\right)  =0$
and $t\left(  \pi\right)  =1$. Then, the length $\mathcal{L}\left[
\gamma\left(  \xi\right)  \right]  _{0\leq\xi\leq\pi}=\mathcal{L}\left[
\gamma\left(  \xi\left(  t\right)  \right)  \right]  _{0\leq t\leq1}$ of the
path $\gamma\left(  \xi\right)  :\xi\mapsto\left\vert \psi\left(  \xi\right)
\right\rangle $ with $0\leq\xi\leq\pi$ equals%
\begin{equation}
\mathcal{L}\left[  \gamma\left(  \xi\right)  \right]  _{0\leq\xi\leq\pi}%
=\int_{0}^{\pi}\sqrt{g_{\mathrm{FS}}\left(  \xi\right)  }d\xi=2\arccos\left[
\left\vert \left\langle A\left\vert B\right.  \right\rangle \right\vert
\right]  \text{,} \label{sara}%
\end{equation}
where $g_{\mathrm{FS}}\left(  \xi\right)  $ in Eq. (\ref{sara}) can be
evaluated by inserting Eq. (\ref{carluccio}) into Eq. (\ref{fubini}). This
concludes our discussion regarding how to reach Eq. (\ref{lunghezza}). For
further technical specifications, we recommend referring to Ref.
\cite{cafarocqg}.

\bigskip


\begin{thebibliography}{99}                                                                                               %


\bibitem {farhi98}E. Farhi and S. Gutmann, \emph{An analog analogue of a
digital quantum computation}, Phys. Rev. \textbf{A57}, 2403 (1998).

\bibitem {carloijqi}C. Cafaro and P. M. Alsing, \emph{Continuous-time quantum
search and time-dependent two-level quantum systems}, Int. J. Quantum
Information \textbf{17}, 1950025 (2019).

\bibitem {steven20}S. Gassner, C. Cafaro, and S. Capozziello, \emph{Transition
probabilities in generalized quantum search Hamiltonian evolutions}, Int. J.
Geom. Meth. Mod. Phys. \textbf{17}, 2050006 (2020).

\bibitem {ali09}A. Mostafazadeh, \emph{Hamiltonians generating optimal-speed
evolutions}, Phys. Rev. \textbf{A79}, 014101 (2009).

\bibitem {bender07}C. M. Bender, D. C. Brody, H. F. Jones, and B. K. Meister,
\emph{Faster than Hermitian quantum mechanics}, Phys. Rev. Lett. \textbf{98},
040403 (2007).

\bibitem {cafarocqg}C. Cafaro and P. M. Alsing, \emph{Qubit geodesics on the
Bloch sphere from optimal-speed Hamiltonian evolutions}, Class. Quantum Grav.
\textbf{40}, 115005 (2023).

\bibitem {brody03}D. C. Brody, \emph{Elementary derivation for passage times},
J. Phys.: Math. Gen. \textbf{36}, 5587 (2003).

\bibitem {grover}L. K. Grover, \emph{Quantum mechanics helps in searching for
a needle in a haystack}, Phys. Rev. Lett. \textbf{79}, 325 (1997).

\bibitem {grover05}L. K. Grover, \emph{Fixed-point quantum search}, Phys. Rev.
Lett. \textbf{95}, 150501 (2005).

\bibitem {zalka}C. Zalka, \emph{Grover's quantum searching algorithm is
optimal}, Phys. Rev. \textbf{A60}, 2746 (1999).

\bibitem {wadati}A. Miyake and M. Wadati, \emph{Geometric strategy for the
optimal quantum search}, Phys. Rev. \textbf{A64}, 042317 (2001).

\bibitem {cafaro17}C. Cafaro, \emph{Geometric algebra and information geometry
for quantum computational software}, Physica \textbf{A470}, 154 (2017).

\bibitem {bae02}J. Bae and Y. Kwon, \emph{Generalized quantum search
Hamiltonians}, Phys. Rev. \textbf{A66}, 012314 (2002).

\bibitem {farhi00}E. Farhi, J. Goldstone, S. Gutmann, and M. Sipser,
\emph{Quantum computation by adiabatic evolution}, arXiv:quant-ph/0001106 (2000).

\bibitem {roland02}J. Roland and N. J. Cerf, \emph{Quantum search by local
adiabatic evolution}, Phys. Rev. \textbf{A65}, 042308 (2002).

\bibitem {sanders04}K.-P. Marzlin and B. C. Sanders, \emph{Inconsistency in
the application of the adiabatic theorem}, Phys. Rev. Lett. \textbf{93},
160408 (2004).

\bibitem {oh05}D. M. Tong, K. Singh, L. C. Kwek, and C. H. Oh,
\emph{Quantitative conditions do not guarantee the validity of the adiabatic
approximation}, Phys. Rev. Lett. \textbf{95}, 110407 (2005).

\bibitem {ali04}M. Andrecut and M. K.\ Ali, \emph{The adiabatic analogue of
the Margolus-Levitin theorem}, J. Phys. A: Math. Gen. \textbf{37}, L157 (2004).

\bibitem {romanelli07}A. Perez and A. Romanelli, \emph{Nonadiabatic quantum
search algorithms}, Phys. Rev. \textbf{A76}, 052318 (2007).

\bibitem {cafaro12a}C. Cafaro and S. Mancini, \emph{An information geometric
viewpoint of algorithms in quantum computing}, in Bayesian Inference and
Maximum Entropy Methods in Science and Engineering, AIP Conf. Proc.
\textbf{1443}, 374 (2012).

\bibitem {cafaro12b}C. Cafaro and S. Mancini, \emph{On Grover's search
algorithm from a quantum information geometry viewpoint}, Physica
\textbf{A391}, 1610 (2012).

\bibitem {anandan90}J. Anandan and Y. Aharonov, \emph{Geometry of quantum
evolution}, Phys. Rev. Lett. \textbf{65}, 1697 (1990).

\bibitem {hamilton23}G. A. Hamilton and B. K. Clark, \emph{Quantifying unitary
flow efficiency and entanglement for many-body localization}, Phys. Rev.
\textbf{B107}, 064203 (2023).

\bibitem {cafaro22}C. Cafaro, S. Ray, and P. M. Alsing, \emph{Optimal-speed
unitary quantum time evolutions and propagation of light with maximal degree
of coherence}, Phys. Rev. \textbf{A105}, 052425 (2022).

\bibitem {wolf59}E. Wolf, \emph{Coherence properties of partially polarized
electromagnetic radiation}, Il Nuovo Cimento \textbf{13}, 1180 (1959).

\bibitem {wolf07}E. Wolf,\emph{ Introduction to the Theory of Coherence and
Polarization of Light}, Cambridge University Press (2007).

\bibitem {rossetti24}L. Rossetti, C. Cafaro, and N. Bahreyni,
\emph{Constructions of optimal-speed quantum evolutions: A comparative study},
Physica Scripta \textbf{99}, 095121 (2024).

\bibitem {haywood}S. Haywood, \emph{Symmetries and Conservation Laws in
Particle Physics}, Imperial College Press (2010).

\bibitem {lu16}D. Lu et \textit{al}., \emph{Chiral quantum walks}, Phys. Rev.
\textbf{A93}, 042302 (2016).

\bibitem {chen17}T. Chen, B.\ Wang, and X. Zhang, \emph{Controlling
probability transfer in the discrete-time quantum walk by modulating the
symmetries}, New J. Phys. \textbf{19}, 113049 (2017).

\bibitem {manzano18}D. Manzano and P. I. Hurtado, \emph{Harnessing symmetry to
control quantum transport}, Adv. Phys. \textbf{67}, 1 (2018).

\bibitem {bottarelli25}A. Bottarelli et \textit{al}., \emph{Symmetry-enhanced
counteradiabatic quantum algorithms for qudits}, Phys. Rev. Research
\textbf{7}, 043030 (2025).

\bibitem {fenner2000}S. Fenner, \emph{An intuitive Hamiltonian for quantum
search}, arXiv:quant-ph/0004091 (2000).

\bibitem {carloepjplus}C. Cafaro, E. Clements, and A. Alanazi, \emph{Aspects
of complexity in quantum evolutions on the Bloch sphere}, Eur. Phys. J. Plus
\textbf{140}, 349 (2025).

\bibitem {uzdin12}R. Uzdin, U. G\"{u}nther, S. Rahav, and N. Moiseyev,
\emph{Time-dependent Hamiltonians with 100\% evolution speed efficiency}, J.
Phys. A: Math. Theor. \textbf{45}, 415304 (2012).

\bibitem {cafaropra20}C. Cafaro, S. Ray, and P. M. Alsing, \emph{Geometric
aspects of analog quantum search evolutions}, Phys. Rev. \textbf{A102, }052607 (2020).

\bibitem {rossetti25}L. Rossetti, C. Cafaro, and P. M. Alsing,
\emph{Deviations from geodesic evolutions and energy waste on the Bloch
sphere}, Phys. Rev. \textbf{A111}, 022441 (2025).

\bibitem {feynman57}R. P. Feynman, F. Vernon, and R. W. Hellwarth,
\emph{Geometrical representation of the Schr\"{o}dinger equation for solving
maser problems}, J. Appl. Phys. \textbf{28}, 49 (1957).

\bibitem {cafaropra25}C. Cafaro, L. Rossetti, and P. M. Alsing,
\emph{Curvature of quantum evolutions for qubits in time-dependent magnetic
fields}, Phys. Rev. \textbf{A111}, 012408 (2025).

\bibitem {donald}D. A. McQuarrie, \emph{Quantum Chemistry}, University Science
Books (2008).

\bibitem {morton}M. Hamermesh, \emph{Group Theory and Its Application to
Physical Problems}, Dover Publications (1989).

\bibitem {bertolucci89}D. C. Harris and M. D. Bertolucci, \emph{Symmetry and
Spectroscopy}, Dover Publications, Inc., New York (1989).

\bibitem {albertini18}F. Albertini and D. D'Alessandro, \emph{On symmetries in
time optimal control, sub-Riemannian geometries, and the K-P problem}, J. Dyn.
Control Sys. \textbf{24}, 13 (2018).

\bibitem {wang21}Y. Wang and S. Wu, \emph{Role of symmetry in quantum search
via continuous-time quantum walk}, SPIN \textbf{11}, 2140002 (2021).

\bibitem {meyer25}D. A. Meyer and T. G. Wang, \emph{Conserved quantities in
linear and nonlinear quantum search}, Quantum Inf. Comput. \textbf{25}, 315 (2025).

\bibitem {popescu24}R. A. Gutoiu, A. Tanasescu, and P. G. Popescu,
\emph{Simple exact quantum search}, Quantum Inf. Process. \textbf{23}, 356 (2024).

\bibitem {wang25}S.-M. Ye and Y.-L. Wang, \emph{Deterministic quantum partial
search for target states with proportion }$1/16$, Phys. Lett. \textbf{A537},
130325 (2025).

\bibitem {nielsen10}M. A. Nielsen and I. L. Chuang, \emph{Quantum Computation
and Quantum Information}, Cambridge University Press (2010).

\bibitem {provost80}J. P. Provost and G. Vallee, \emph{Riemannian structure on
manifolds of quantum states}, Commun. Math. Phys. \textbf{76}, 289 (1980).

\bibitem {brau94}S. Braunstein and C. M. Caves, \emph{Statistical distance and
the geometry of quantum states}, Phys. Rev. Lett. \textbf{72}, 3439 (1994).
\end{thebibliography}
\end{document}